\documentclass[sigconf]{acmart}

\copyrightyear{2026}
\acmYear{2026}
\setcopyright{cc}
\setcctype{by}
\acmConference[CHI '26]{Proceedings of the 2026 CHI Conference on Human Factors in Computing Systems}{April 13--17, 2026}{Barcelona, Spain}
\acmBooktitle{Proceedings of the 2026 CHI Conference on Human Factors in Computing Systems (CHI '26), April 13--17, 2026, Barcelona, Spain}
\acmPrice{}
\acmDOI{10.1145/3772318.3790339}
\acmISBN{979-8-4007-2278-3/2026/04}

\usepackage{booktabs} 
\usepackage{array} 
\usepackage{caption} 
\begin{document}
\title[Large Language Models in Peer-Run Community Behavioral Health Services]{Large Language Models in Peer-Run Community Behavioral Health Services: Understanding Peer Specialists and Service Users’ Perspectives on Opportunities, Risks, and Mitigation Strategies}

\author{Cindy Peng}
%\authornote{Corresponding author.}
\affiliation{%
  \department{School of Computer Science}
  \institution{Carnegie Mellon University}
  \city{Pittsburgh}
  \state{Pennsylvania}
  \country{USA}
}
\email{cindypen@andrew.cmu.edu}

\author{Megan Chai}
\affiliation{%
  \department{Human-Computer Interaction Institute}
  \institution{Carnegie Mellon University}
  \city{Pittsburgh}
  \state{Pennsylvania}
  \country{USA}
}
\email{mvchai@andrew.cmu.edu}

\author{Gao Mo}
\affiliation{%
  \department{School of Computer Science}
  \institution{Carnegie Mellon University}
  \city{Pittsburgh}
  \state{Pennsylvania}
  \country{USA}
}
\email{gaom@cs.cmu.edu}

\author{Naveen Raman}
\affiliation{%
  \department{School of Computer Science}
  \institution{Carnegie Mellon University}
  \city{Pittsburgh}
  \state{Pennsylvania}
  \country{USA}
}
\email{naveenr@cmu.edu}

\author{Ningjing Tang}
\affiliation{%
  \department{Human-Computer Interaction Institute}
  \institution{Carnegie Mellon University}
  \city{Pittsburgh}
  \state{Pennsylvania}
  \country{USA}
}
\email{ningjint@andrew.cmu.edu}

\author{Shannon Pagdon}
\affiliation{%
  \department{School of Social Work}
  \institution{University of Pittsburgh}
  \city{Pittsburgh}
  \state{Pennsylvania}
  \country{USA}
}
\email{shp121@pitt.edu}

\author{Margaret Swarbrick}
\affiliation{%
  \department{Graduate School of Applied and Professional Psychology}
  \institution{Rutgers University}
  \city{Piscataway}
  \state{New Jersey}
  \country{USA}
}
\email{swarbrma@rutgers.edu}

\author{Nev Jones}
\affiliation{%
  \department{School of Social Work}
  \institution{University of Pittsburgh}
  \city{Pittsburgh}
  \state{Pennsylvania}
  \country{USA}
}
\email{nevjones@pitt.edu}

\author{Fei Fang}
\affiliation{%
  \department{Software and Societal Systems Department}
  \institution{Carnegie Mellon University}
  \city{Pittsburgh}
  \state{Pennsylvania}
  \country{USA}
}
\email{feifang@cmu.edu}

\author{Hong Shen}
\affiliation{%
  \department{Human-Computer Interaction Institute}
  \institution{Carnegie Mellon University}
  \city{Pittsburgh}
  \state{Pennsylvania}
  \country{USA}
}
\email{hongs@cs.cmu.edu}

\renewcommand{\shortauthors}{Peng et al.}

\begin{abstract}
    Peer-run organizations (PROs) provide critical, recovery-based behavioral health support rooted in lived experience. As large language models (LLMs) enter this domain, their scale, conversationality, and opacity introduce new challenges for situatedness, trust, and autonomy. Partnering with Collaborative Support Programs of New Jersey (CSPNJ), a statewide PRO in the Northeastern United States, we used comicboarding, a co-design method, to conduct workshops with 16 peer specialists and 10 service users exploring perceptions of integrating an LLM-based recommendation system into peer support. Findings show that depending on how LLMs are introduced, constrained, and co-used, they can reconfigure in-room dynamics by sustaining, undermining, or amplifying the relational authority that grounds peer support. We identify opportunities, risks, and mitigation strategies across three tensions: bridging scale and locality, protecting trust and relational dynamics, and preserving peer autonomy amid efficiency gains. We contribute design implications that center lived-experience-in-the-loop, reframe trust as co-constructed, and position LLMs not as clinical tools but as relational collaborators in high-stakes, community-led care.
\end{abstract}

%old version
%\begin{CCSXML}
%<ccs2012>
%<concept>
%<concept_id>10003120.10003121.10003122.10003334</concept_id>
%<concept_desc>Human-centered computing~User studies</concept_desc>
%<concept_significance>500</concept_significance>
%</concept>
%<concept>
%<concept_id>10003120.10003121.10011748</concept_id>
%<concept_desc>Human-centered computing~Empirical studies in HCI</concept_desc>
%<concept_significance>500</concept_significance>
%</concept>
%</ccs2012>
%\end{CCSXML}

%\ccsdesc[500]{Human-centered computing~User studies}
%\ccsdesc[500]{Human-centered computing~Empirical studies in HCI}

%new version



  %\includegraphics[width=\textwidth]{images/Cover.pdf}
  %\caption{Comicboarding co-design method used to ground conversations with peer specialists and service users about integrating LLMs into peer-run behavioral health settings, prompting discussion of opportunities, risks, and mitigation strategies for everyday peer support.}
  %\Description{A example of a set of comicboards that we showed participants that consists of five panels, each of a visual image depicting the potential LLM system design or implementation or prompting a question to generate ideas about its design.}
  %\label{fig:teaser}



%   \includegraphics[width=\textwidth]{teaser-figure}

%\received{September 11, 2025} %date of original submission
%\received[revised]{December 04, 2025} %date of revisions
%\received[accepted]{January 15, 2026} %date of acceptance


\maketitle

\section{Introduction}

In the United States, the number of people with behavioral health conditions continues to outpace access to quality treatment and care \cite{mechanic2014more}. Undiagnosed and undertreated conditions are a leading source of disability, reduced quality of life, and premature mortality \cite{velligan2023epidemiology, fiorillo2021mortality}. The resulting treatment gap imposes substantial social and economic burdens \cite{thorpe2017prevalence, taylor2023economic}. For example, individuals with serious behavioral health conditions such as schizophrenia experience substantially reduced life expectancy, often 15 to 20 years shorter than the general population, and face extremely high rates of unemployment or underemployment, between 70\% and 90\% \cite{peritogiannis2022mortality, franke2024association, kadakia2022economic}. These disparities underscore the urgency of developing more accessible, community-centered care solutions.
 
Peer-run organizations (PROs) play a vital role in addressing behavioral health disparities by offering non-clinical, community-based support led by individuals with lived experience of mental health conditions or substance use disorders \cite{ostrow2015leadership, simmons2023effectiveness}. They adopt a distinct, holistic philosophy of wellness grounded in the Eight Dimensions of Wellness model \cite{swarbrick2012wellness, nemec2021nudges, swarbrick2025factor}, a strengths-based framework that defines wellbeing across emotional, social, physical, intellectual, spiritual, occupational, financial, and environmental domains rather than diagnosis or treatment alone \cite{swarbrick2006wellness}. Widely adopted by PROs and community-based interventions \cite{mallahan2023qualitative, bellamy2021collaborative, jordan2023breaking}, this model conceptualizes wellness as dynamic, nonlinear, and shaped by interconnected life contexts. By emphasizing personal agency, self-direction, and sustainable habits, PROs help individuals build routines that promote stability, resilience, and long-term wellbeing through a whole-person, lifestyle-oriented approach \cite{swarbrick2025factor}. The framework centers empowerment, mutual growth, and community connection, honoring lived experience, dignity, and socioeconomic realities throughout the recovery journey \cite{swarbrick2021reciprocal}. PROs often support individuals with multifaceted social needs, including navigating homelessness, co-occurring conditions, and intersecting structural disadvantages, who are frequently described by institutions as ``difficult to reach and engage'' due to compounding experiences of marginalization and exclusion \cite{ostrow2015leadership, mangan2024peer, kirkbride2024social}. Beyond behavioral health services, PROs address broader systemic barriers such as social stigma and poverty, and connect people to critical resources, including housing assistance, legal advocacy, and financial empowerment \cite{ostrow2015leadership,wolf2024changing,swarbrick2007consumer,tanenbaum2011mental}.

Despite their essential contribution, PROs remain largely under-resourced and face persistent systemic challenges such as inadequate funding, limited technological infrastructure, and growing demand \cite{blash2015peer, ostrow2017medicaid, ostrow2015leadership}. Many rely on in-person, paper-based workflows and lack the capacity to adopt or maintain digital tools \cite{campbell1997data, ostrow2014improving, segal2016consumer}. These constraints hinder service quality and scalability, even as the need for comprehensive behavioral health support continues to grow.

At the same time, artificial intelligence (AI)---particularly large language models (LLMs)---is rapidly entering behavioral health settings \cite{xu2024mental, lee2024influence, lai2023psy} due to its ability to generate natural language, synthesize information, and support complex, context-rich tasks that align closely with everyday peer support work. Yet many of these systems are developed without meaningful input from the communities they aim to serve and are often evaluated only after deployment \cite{chen2019can,fiske2019your,gooding2021ethics}. They may also introduce safety risks, such as hallucinated or overconfident outputs \cite{hallucination_llm}. These concerns make their use particularly consequential in high-stakes, trauma-informed contexts and warrant careful consideration in behavioral health settings.

While prior work in the human-computer interaction (HCI) community has made significant strides in promoting community-centered approaches to AI design \cite{katell2020toward,robertson2021modeling,lee2019webuildai,zhu2018value,krafft2021action,brown2019toward,eslami2025margins}, extending these practices to LLMs in peer-run behavioral health settings presents new challenges. LLMs’ expansive capabilities and difficult-to-predict behaviors complicate stakeholder comprehension and early-stage ideation \cite{llm_blackbox}. Moreover, the communities served by PROs often face intersecting structural barriers, have limited access to technology, and may feel unqualified to shape system design processes \cite{kuo2023understanding,tang2024ai}. Despite their central roles, frontline peer specialists and service users---two key stakeholder groups---are rarely involved in the design of LLM-based tools, and little is known about how they perceive such systems or envision their impact on peer support.

To address this gap, we partnered with Collaborative Support Programs of New Jersey (CSPNJ), a statewide, award-winning PRO in the Northeastern United States, to investigate how peer specialists and service users perceive the integration of an LLM-based recommendation system into their everyday workflows (see Section~\ref{sec:studycontext}). We ask: 
\begin{itemize}
    \item \textbf{\textit{RQ1:} }How do peer specialists and service users perceive the potential opportunities of integrating LLMs into peer support?
    \item \textbf{\textit{RQ2:} }How do they perceive the potential risks of integrating LLMs into peer support?
    \item \textbf{\textit{RQ3:} }How do they envision strategies to mitigate the potential risks associated with integrating LLMs into peer support?
\end{itemize}

To answer these questions, we conducted workshops with 16 peer specialists and 10 service users using comicboarding \cite{moraveji2007comicboarding}, a co-design method to elicit feedback on LLM capabilities, limitations, and improvement pathways (see Section~\ref{sec:llmboards}, Section~\ref{sec:protocol}, and Section~\ref{sec:participants}). This approach enabled rich, grounded conversations with participants whose lived expertise is often overlooked in behavioral health AI development.

Our findings surface one central theme: LLMs can shift the balance of relational authority in peer support depending on how they are used, interacted with, and implemented (see Section~\ref{sec:results}). This theme manifests through three tensions shaping how LLMs may be responsibly integrated into peer-run behavioral health services. First, participants emphasized the need to bridge scale-trained outputs with localized, experiential wisdom, noting that generic LLM responses often misalign with the situated knowledge and cultural nuance required in peer support. Second, they highlighted risks to relational trust, describing how LLMs could inadvertently disrupt the co-constructed, trauma-informed dynamics that define effective peer-user relationships. Third, they raised concerns about peer autonomy, warning that automation may displace the authority of lived experience and deskill peer workers over time. Building on these insights, we propose design implications that center ``lived-experience-in-the-loop,'' reframe trust as co-constructed rather than transferred, and reposition LLMs not as clinical decision-makers but as relational collaborators in high-stakes, community-led care (see Section~\ref{sec:discussion}). These implications extend beyond peer support to inform the design of LLM systems in other settings where care, power, and expertise are negotiated through social interaction.

In this study, our contributions are threefold:
\begin{enumerate}
    \item We offer \textit{\textbf{empirical insights}} into how peer specialists and service users---frontline stakeholders in behavioral health---perceive the opportunities, risks, and mitigation strategies of integrating LLMs into peer support workflows.
    \item We introduce \textbf{\textit{lived-experience-in-the-loop as a design principle}} for LLM-supported care, reframing judgment as experiential, trust as co-constructed, and authority as relational in LLM-human collaboration in high-stakes support contexts.
    \item We derive \textbf{\textit{actionable design implications}} for the responsible and context-sensitive integration of LLMs in relational, community-led services grounded in local context, trust-building practices, and mechanisms for repair and autonomy preservation.
\end{enumerate}
\section{Related Work}

\subsection{Peer-Run Organizations in Community-Based Behavioral Health Services}
\label{pros}
Peer-run organizations (PROs) deliver community-based behavioral health services to individuals with mental health conditions and are led by peer specialists who self-identify as having lived experience with such conditions \cite{swarbrick2025lived, wolf2024changing}. Amid a worsening behavioral health workforce crisis, policymakers and researchers have increasingly advocated shifting tasks from professional to paraprofessional providers \cite{gaiser2021systematic, hoeft2018task, ostrow2024employment}. Within PROs and public-sector behavioral health services, peer specialists fulfill a variety of roles, including providing emotional, social, and practical support grounded in shared lived experience \cite{swarbrick2025lived, solomon2004peer}. Through positive self-disclosure and role modeling, peer specialists play a crucial role in PROs \cite{davidson2012peer}. Rather than approaching service users through a traditional clinical lens, they often engage bidirectionally with service users, sharing experiences and building relationships through mutual empowerment outside a clinical framework \cite{penney2021development}. Fostering strong relationships can increase service users’ sense of hope, self-efficacy, and self-worth within a community united by a shared focus on recovery \cite{davidson2012peer, alberta2012addressing, wolf2024changing}, which the Substance Abuse and Mental Health Services Administration (SAMHSA) defines as: \textit{``A process of change through which individuals improve their health and wellness, live a self-directed life, and strive to reach their full potential''} \cite{substance2012samhsa}. 

Historically rooted in the consumer/survivor/ex-patient movement, PROs provide a wide range of critical safety-net services, including resource referral, housing and employment support, mutual aid, legal advocacy, and peer counseling, to support community integration for individuals with behavioral health challenges \cite{swarbrick2007consumer, tanenbaum2011mental, wolf2024changing}. Unlike more conventional clinical services, recovery-oriented peer support emphasizes holistic healing. Beyond psychological and behavioral health support, peer specialists also offer housing assistance, benefits management, financial empowerment, community outreach, systems navigation, and advocacy \cite{ostrow2015leadership}. Research further suggests that PROs provide stronger support for the peer specialists they employ, including increased opportunities for career mobility and advancement and a more supportive organizational climate \cite{jones2020organizational}. 

Despite their growing prominence, PROs remain underfunded and technologically under-resourced within the behavioral health sector \cite{blash2015peer, ostrow2017medicaid, ostrow2015leadership}. Many lack the infrastructure needed to support health technologies, electronic health record (EHR) management, and data-driven quality evaluation \cite{campbell1997data, ostrow2014improving, segal2016consumer}, which limits their ability to track and manage the types and range of wellness support provided to the people they serve. As a result, many PROs rely on in-person, paper-based workflows, making it difficult to meet rising demand and address challenges faced by both peer specialists and the communities they support. As increasingly complex tasks are shifted to the peer sector and caseloads grow, peer specialists may struggle to meet the intersecting needs they encounter, such as poverty, food insecurity, high rates of physical comorbidities (e.g., diabetes and HIV), substance use, and psychiatric disability \cite{smith2023food, walker2017cumulative, draine2013mental, alberta2012addressing}. These compounding pressures underscore the potential value and appeal of leveraging intelligent, data-driven tools such as LLMs to improve peer workflows, expand resource navigation, and enhance recovery-oriented service delivery. In this study, we address this gap by examining how LLMs can be introduced within the real constraints and values of CSPNJ (see Section~\ref{sec:studycontext}), identifying where such tools can meaningfully extend organizational capacity and where they must defer to peer judgment to preserve relationship-centered practice.

\subsection{LLMs in Behavioral Health Contexts and Practice}
\label{llms}
Large language models (LLMs) are a class of AI technology that leverage large-scale data to mimic human language by scaling model parameters to tens or hundreds of billions and utilizing extensive training corpora \cite{naveed2023comprehensive}. Due to their generality, LLMs have been applied to a wide range of complex natural language processing (NLP) tasks that approximate human-level performance, including machine translation \cite{lyu2023paradigm}, summarization \cite{liu2023learning}, and text generation \cite{chung2023increasing}. This rapid innovation has transformed how AI systems are designed and deployed, with applications such as ChatGPT illustrating their broad societal impact \cite{zhao2023survey}. In recent years, LLM-powered systems have been increasingly implemented in behavioral health contexts, including mental health prediction \cite{xu2024mental}, mental health intervention chatbots \cite{lee2024influence}, and psychological consultation \cite{lai2023psy}. Because many of these systems have been developed without community engagement, our work aims to bridge the gap between behavioral health service users, frontline providers, and LLM developers \cite{gooding2021ethics} by incorporating the perspectives and lived experiences of these communities.

The unreliability of LLMs presents a major limitation when applying them to real-world settings \cite{systematic_llm_reliability}. LLMs can generate hallucinations, referring to instances in which the model produces factually incorrect information \cite{hallucination_llm}. The risks associated with hallucinations are particularly salient in high-stakes domains. For example, hallucinated outputs may be inappropriate \cite{jo2023understanding} or lead to harmful outcomes for the populations being served \cite{guo2024large}. Furthermore, the black-box nature of LLMs makes it difficult to understand \textit{why} hallucinations occur \cite{guo2024large}. Despite the widespread adoption of LLM-powered applications in behavioral health and the well-documented risks, many systems continue to be developed without meaningful engagement with the communities they aim to serve \cite{chan2021limits}. This \textit{``acoustic separation''} can result in ineffective outcomes or even harmful consequences in high-stakes, trauma-informed domains, disproportionately affecting disadvantaged social groups relying on public behavioral health services \cite{chen2019can,fiske2019your,gooding2021ethics}. Furthermore, much LLM research in behavioral health centers clinicians, institutions, or general-purpose users rather than peer-run services, leaving a gap in understanding how LLMs intersect with recovery-oriented peer practice. 

\textit{In this study, we do not assume that LLMs are appropriate or inevitable in peer-run care contexts.} Rather, in light of the rapid adoption of these technologies, our goal is to foreground the perspectives of those most directly affected---peer specialists and service users---and to examine how LLMs might be designed, constrained, or resisted to support, rather than undermine, community-led behavioral health practice. With CSPNJ interested in exploring the integration of an LLM-powered recommendation system (see Section~\ref{sec:studycontext}), we investigate the context-specific opportunities, risks, and mitigation strategies associated with introducing LLMs into peer-focused behavioral health service delivery, with the aim of informing relationship-centered, community-grounded design.

\subsection{Community-Centered AI Design with Peer-Run Community Stakeholders}
\label{community-centered}

With the growing integration of AI and LLMs into real-world communities, there have been increasing calls within the HCI and CHI communities to re-center impacted stakeholders---those at the receiving end of AI decisions and most likely to be harmed---in the design of the systems deployed in their communities \cite{katell2020toward, robertson2021modeling, lee2019webuildai, zhu2018value, krafft2021action, brown2019toward, kuo2023understanding}. Past research has contributed to important insights and empirical lessons on how to align AI systems with the values, needs, and practices of affected communities \cite{lee2024ai, wang2022co, kawakami2022care, li2024re,tseng2025ownership}. For example, participatory design approaches have been used to surface children’s understanding of AI’s influence on culture \cite{dangol2024mediating}. Other studies have examined AI and LLM use in clinical and public health settings \cite{yang2023large}, including work that co-designs AI decision-support tools with healthcare practitioners \cite{hussain2024development}. 

Designing community-centered LLMs for peer-run behavioral health services presents unique challenges. On the one hand, LLMs are complex, opaque systems with broad capabilities and emergent behaviors, making them difficult to understand or critique, even for technical experts \cite{llm_blackbox}. On the other hand, for peer specialists and service users within PROs, these technical hurdles are compounded by intersecting structural disadvantages. Many service users live with serious health conditions alongside housing instability, poverty, stigma, and low literacy, all of which further limit participation in AI design processes \cite{saatcioglu2014poverty,jenkins2023social, karadzhov2020coping, kaite2015ongoing, tweed2021health}. These conditions restrict digital access, constrain opportunities for civic or design participation, and contribute to broader patterns of epistemic exclusion. As a result, these communities---despite being among those most impacted by behavioral health innovations---are routinely overlooked in AI development processes that privilege more resourced and technically literate perspectives \cite{donia2021co, zidaru2021ensuring}.

To support the meaningful inclusion of peer specialists and service users in LLM design, our study focuses on expanding access to design processes for communities historically excluded from shaping the technologies that affect their lives. Prior HCI research has developed a range of methods for engaging diverse stakeholders in AI ideation and evaluation \cite{brown2019toward, lee2019webuildai, zhu2018value}, including identifying failure points \cite{tang2024ai}, co-evaluating policies \cite{krafft2021action}, and aligning system explanations with end-user values \cite{lee2024ai}. However, these approaches have rarely centered LLMs within peer-run public behavioral health contexts. Building on this foundation, we adapt comicboarding (see Section~\ref{sec:llmboards}) that lowers participation barriers for individuals with varying levels of technical literacy \cite{moraveji2007comicboarding, hiniker2017co, lee2021show}. This approach enabled us to elicit concrete feedback from frontline peer specialists and service users, including individuals navigating some of the most profound structural disadvantages in our society, and to concretize LLM implications in peer support practice through design-ready insights grounded in PRO workflows and informed by community stakeholders.

\section{Study Design}

\begin{table*}[t]
\centering
\caption{Key steps in peer-led resource navigation and support workflows, illustrating common practices used by peer specialists to support service users at CSPNJ community wellness centers.}
\Description{A two-column table listing key steps in peer-led resource navigation workflows. The left column names each step, including clarifying the resource need, locating appropriate resources, assessing fit and accessibility, supporting application and connection tasks, skill-building for ongoing access, and follow-up and troubleshooting. The right column lists bullet-point examples of actions peer specialists take at each step to support service users.}
\label{tab:peer-workflows}
%\footnotesize
\renewcommand{\arraystretch}{1.3}
\begin{tabular}{p{7cm} p{10.0cm}}
\toprule
\textbf{Clarifying the Resource Need} & 
\textbullet\ Translate the service user’s concerns into specific resource needs (e.g., food assistance, transportation, benefits, healthcare, legal aid).

\textbullet\ Identify urgency, eligibility considerations, and potential barriers (e.g., documentation, technology access, time constraints). 
\\
\midrule
\textbf{Locating Appropriate Resources} &
\textbullet\ Search internal and community resource directories.

\textbullet\ Identify culturally relevant options.

\textbullet\ Compare multiple resources, when available, to support informed choice.
\\
\midrule
\textbf{Assessing Fit and Accessibility} &
\textbullet\ Review eligibility criteria, cost, location, hours, language access, and referral requirements.

\textbullet\ Discuss the pros and cons of each option with the service user. 

\textbullet\ Flag potential barriers (e.g., waitlists, transportation, or digital access). 
\\
\midrule
\textbf{Supporting Application and Connection Tasks} &
\textbullet\ Assist the service user in completing forms or online applications without taking control.

\textbullet\ Support the gathering of required information or documentation.  

\textbullet\ Help schedule appointments or plan next steps.
\\
\midrule
\textbf{Skill-Building for Ongoing Access} &
\textbullet\ Model how to search for resources independently. 

\textbullet\ Practice scripts for self-advocacy and follow-up.  

\textbullet\ Share strategies for tracking applications, appointments, or deadlines. 
\\
\midrule
\textbf{Follow-Up and Troubleshooting} &
\textbullet\ Check in with the service user on whether the resource connection was successful. 

\textbullet\ Problem-solve barriers or denials.  

\textbullet\ Identify alternative options if initial attempts were unsuccessful. 
\\
\bottomrule
\end{tabular}
\end{table*}

\subsection{Study Context}
\label{sec:studycontext}
This study was part of a larger research initiative conducted in collaboration with Collaborative Support Programs of New Jersey (CSPNJ), a statewide, peer-run behavioral health organization based in New Jersey, United States. Incorporated in 1984, CSPNJ operates across more than 30 locations and is nationally recognized for its leadership in the design and delivery of wellness- and recovery-oriented services. Known for its award-winning work, CSPNJ has decades of experience supporting individuals with multifaceted social needs, including those navigating mental health challenges, substance use concerns, housing instability, and behavioral health crises. 

As a peer-run agency, CSPNJ’s services are staffed and led by people with lived experience of recovery from similar challenges, positioning peer specialists to cultivate trust, empathy, and egalitarian forms of support that differ from traditional clinical roles. CSPNJ’s programs emphasize holistic wellness across multiple dimensions, creating opportunities for education, skill-building, crisis management, access to food and housing, recovery peer sessions, and connection to community resources. Across its network of 15 community wellness centers, CSPNJ serves between 15 and 100 service users per day. Annually, it reaches approximately 675 individuals through supportive housing programs and around 700 through peer wellness respites. 

In practice, peer support interactions at CSPNJ are shaped by collaborative, relationship-centered workflows that foreground service user agency (see Table~\ref{tab:peer-workflows}). A typical encounter begins with welcoming the service user and establishing rapport through active listening in a respectful, non-judgmental space. Peer specialists then work with the service user to identify current wellness priorities, exploring immediate needs and longer-term goals across areas such as housing, food access, healthcare, financial stability, social connection, or daily routines, and clarifying what support feels most helpful and manageable in the moment. Peer specialists provide relevant information about available resources and options and assist with practical tasks as needed, such as reviewing eligibility requirements, completing forms, locating contact information, role-playing conversations, or supporting calls and appointment scheduling. Throughout the interaction, peer specialists emphasize choice, strengths, and self-direction, supporting service users in deciding which steps to take and when. Encounters often conclude with a brief plan for next steps and, when appropriate, an agreement to follow up, reinforcing continuity of support and progress toward wellness goals. 

Despite this impact, CSPNJ faces systemic constraints common among peer-run providers. Demand for in-person services remains high, while funding and resources to expand support are limited. Technological infrastructure is inadequate, with low-tech tools and a lack of systems for digital service delivery. Peer specialists are frequently expected to multitask without clear role definitions or sufficient technology training. Together, these realities create a compelling case for thoughtfully introduced technologies, including AI. Yet successful integration requires organizational readiness, clear communication, role clarity, and change management. While AI tools hold promise, effective adoption depends on expert-informed design, robust training protocols, and safeguards to ensure safe use.

Our partnership aims to co-develop community-centered AI solutions that address the needs of both peer specialists and service users in resource-constrained public behavioral health contexts. Prior need-finding efforts with CSPNJ identified opportunities and established the potential value of an LLM-based recommendation system to help peer specialists deliver personalized recommendations across services, including but not limited to wellness activities, community resources, and government benefits. However, little is known about how frontline peer specialists and service users perceive the impact of such systems on existing peer-support dynamics, what risks might arise during real-world implementation, and how those risks could be mitigated. This gap motivates the present study.

\begin{figure*}
    \centering
    \includegraphics[width=1\linewidth]{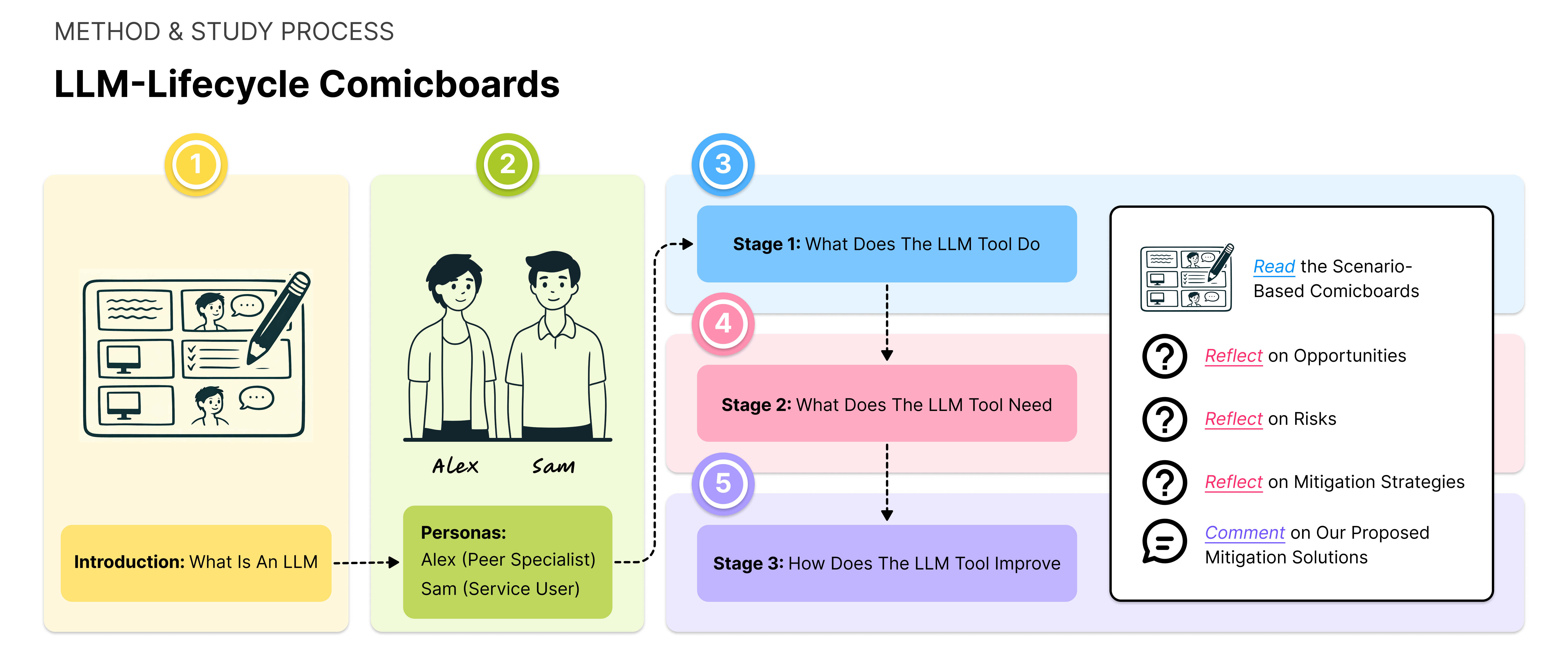}
    \caption{Method and study process using LLM-Lifecycle Comicboards. Participants review an introductory comicboard explaining what an LLM system is (see Appendix Figure~\ref{fig:comicboard0}), familiarize themselves with two personas, then engage with three stages of the LLM lifecycle: (1) what the LLM system can do (see Appendix Figure~\ref{fig:comicboard1}), (2) what it needs (see Appendix Figure~\ref{fig:comicboard2}), and (3) how it gathers feedback to improve (see Appendix Figure~\ref{fig:comicboard3}). At each stage, participants read scenario-based comicboards, reflect on potential opportunities and risks, suggest mitigation strategies, and comment on the research team’s proposed mitigation solutions (see Appendix Figure~\ref{fig:comicboard4}).}
     \Description{A flow diagram illustrating the study process using LLM-lifecycle comicboards. The figure begins with an introductory comicboard explaining what an LLM is, followed by two personas labeled Alex (Peer Specialist) and Sam (Service User). The process then moves through three stages: Stage 1 showing what the LLM tool does, Stage 2 showing what the LLM tool needs, and Stage 3 showing how the LLM tool improves. A side panel lists repeated participant activities at each stage: reading scenario-based comicboards, reflecting on opportunities, risks, and mitigation strategies, then commenting on the research team's proposed mitigation solutions.}
    \label{fig:method_process}
\end{figure*}

\begin{table*}
    \centering
    \caption{Participant self-reported demographic information for peer specialists and service users. Given the sensitive nature of this study, all demographic data are reported in aggregate.}
    \Description{A table reporting participant self-reported demographic information in aggregate. Rows include race, age, gender, years involved in peer support, types of peer support provided or received, and frequency of generative AI use during and outside peer support. Separate columns report counts for peer specialists and service users.}
    \label{tab:participant_table}
    %\footnotesize
    \renewcommand{\arraystretch}{1.3} %newly added
    %\rowcolors{2}{gray!10}{white} %newly added!
    \begin{tabular}{@{}p{9cm}p{4cm}p{4cm}@{}}
    \toprule
         \textbf{Demographic Information}& \textbf{Peer Specialist Counts}& \textbf{Service User Counts}\\ 
         \midrule
         Race &
    \begin{tabular}[t]{@{}p{4cm}@{}}
    \raggedright
    \renewcommand{\arraystretch}{1.25}
    African American (6)\\
    Hispanic (2)\\
    White (8)
    \end{tabular}
    &
    \begin{tabular}[t]{@{}p{4cm}@{}}
    \raggedright
    \renewcommand{\arraystretch}{1.25}
    African American (6)\\
    Hispanic (3)\\
    Asian (1)
    \end{tabular}
    \\
    \midrule

    Age &
    \begin{tabular}[t]{@{}p{4cm}@{}}
    \raggedright
    \renewcommand{\arraystretch}{1.25}
    Mean: 42.6\\
    Minimum: 25\\
    Maximum: 64
    \end{tabular}
    &
    \begin{tabular}[t]{@{}p{4cm}@{}}
    \raggedright
    \renewcommand{\arraystretch}{1.25}
    Mean: 34.3\\
    Minimum: under 25\\
    Maximum: 54
    \end{tabular}
    \\
    \midrule

    Gender &
    \begin{tabular}[t]{@{}p{4cm}@{}}
    \raggedright
    \renewcommand{\arraystretch}{1.25}
    Female (12)\\
    Male (4)
    \end{tabular}
    &
    \begin{tabular}[t]{@{}p{4cm}@{}}
    \raggedright
    \renewcommand{\arraystretch}{1.25}
    Female (3)\\
    Male (7)
    \end{tabular}
    \\
    \midrule

    Years involved &
    Mean: 6.2 &
    Mean: 3.56
    \\
    \midrule

    Peer support provided or received &
    \begin{tabular}[t]{@{}p{4cm}@{}}
    \raggedright
    \renewcommand{\arraystretch}{1.25}
    Crisis intervention\\
    Financial empowerment\\
    Job placement\\
    Listening\\
    Mental health\\
    Resource eligibility\\
    Substance abuse
    \end{tabular}
    &
    \begin{tabular}[t]{@{}p{4cm}@{}}
    \raggedright
    \renewcommand{\arraystretch}{1.25}
    Addiction\\
    Food\\
    General support\\
    Housing\\
    Occupation\\
    Peer services\\
    Resource referrals
    \end{tabular}
    \\
    \midrule

    Frequency of Generative AI (LLM) use during peer support &
    \begin{tabular}[t]{@{}p{4cm}@{}}
    \raggedright
    \renewcommand{\arraystretch}{1.25}
    Very often (3)\\
    Sometimes (3)\\
    Not often (10)
    \end{tabular}
    &
    Never (10)
    \\
    \midrule

    Frequency of Generative AI (LLM) use outside peer support &
    \begin{tabular}[t]{@{}p{4cm}@{}}
    \raggedright
    \renewcommand{\arraystretch}{1.25}
    Very often (4)\\
    Sometimes (6)\\
    Not often (6)
    \end{tabular}
    &
    \begin{tabular}[t]{@{}p{4cm}@{}}
    \raggedright
    \renewcommand{\arraystretch}{1.25}
    Always (1)\\
    Often (1)\\
    Sometimes (2)\\
    Rarely (1)\\
    Never (5)
    \end{tabular}
    \\
        
  \bottomrule

    \end{tabular}
\end{table*}

\subsection{LLM-Lifecycle (Opportunities, Risks, and Mitigation Strategies) Comicboards}
\label{sec:llmboards}
To support meaningful participation from stakeholders with diverse levels of literacy, digital access, and technical familiarity, we adopted comicboarding, a co-design method rooted in HCI research that uses visual storytelling and unfinished comic strips to scaffold participation and support ideation \cite{moraveji2007comicboarding, tang2024ai, kuo2023understanding}. Comicboarding was well suited to our context because peer specialists and service users often encounter literacy barriers, stigma, and other structural challenges that make engagement with abstract or technical concepts such as LLMs difficult \cite{saatcioglu2014poverty, jenkins2023social, karadzhov2020coping, kaite2015ongoing, tweed2021health,hiniker2017co}. Building on prior work engaging communities with multifaceted social needs \cite{kuo2023understanding}, comicboards enabled participation across varied levels of reading and technology literacy \cite{hiniker2017co, moraveji2007comicboarding}.
 
Our approach extends prior comicboarding applications by explicitly structuring boards around key lifecycle stages of an LLM-based recommendation system and its integration into peer-run services (see Figure~\ref{fig:method_process}). We first developed an introductory comicboard that explains \textbf{\textit{(0) what an LLM is}}, grounded in four core aspects of the LLM development process: \textit{pre-training}, \textit{supervised fine-tuning}, \textit{reinforcement learning}, and \textit{prompting} (see Appendix Figure~\ref{fig:comicboard0}) \cite{minaee2024large, yang2024harnessing, naveed2025comprehensive, vaswani2017attention}. We then created three sequential sets of comicboards: \textbf{\textit{(1) what the LLM system can do}} (\textit{capabilities}; see Appendix Figure~\ref{fig:comicboard1}) \cite{zhao2023survey}, \textbf{\textit{(2) what it needs}} (\textit{data and infrastructure}; see Appendix Figure~\ref{fig:comicboard2}) \cite{hoffmann2022training}, and \textbf{\textit{(3) how it learns from feedback}} (\textit{improvement loops}; see Appendix Figure~\ref{fig:comicboard3}) \cite{kim2023aligning}. To ground these scenarios in realistic contexts, we used two personas: a peer specialist and a service user. For each of the three LLM lifecycle stages, the comicboards begin with a sequence of three panels illustrating how LLMs might be integrated into peer support sessions. These visual narratives are followed by three guiding prompts that ask participants to reflect on: \textbf{\textit{(1) potential opportunities}}, \textbf{\textit{(2) potential risks}}, and \textbf{\textit{(3) mitigation strategies}}. We also included \textit{``seed''} examples of risks and mitigations as part of the \textbf{\textit{(4) our mitigation solutions}} comicboards (see Appendix Figure~\ref{fig:comicboard4})---grounded in prior literature and research team brainstorming---to avoid frustration associated with open-ended, \textit{``blue-sky''} ideation, particularly among communities less familiar with design practices \cite{harrington2019deconstructing}. The comicboards were iteratively developed with input from peer specialists within our partner organization, ensuring contextual accuracy and alignment with lived experience. They covered potential LLM applications in wellness planning, resource navigation, and benefits access---areas identified by CSPNJ as especially high need. 

\subsection{Study Protocol}
\label{sec:protocol}
Using these comicboards, we conducted a series of workshops to explore peer specialists’ and service users’ perceptions. We chose workshops over one-on-one interviews because they encourage richer, more collaborative discussion \cite{wilkinson1998focus}. To support open and candid dialogue, we held separate sessions for peer specialists and service users, allowing each group to speak freely within a shared lived-experience context \cite{wilkinson1998focus}. Across workshops, both groups received identical study materials. Before each session, participants completed a self-reported demographic survey and provided informed consent. We began each workshop by using the introductory comicboard to explain the concept of LLMs through the following definition: \textit{A large language model (LLM) is a computer program that understands and generates human-like text based on language patterns it has learned} \cite{minaee2024large}. We then introduced the \textit{\textbf{LLM-Lifecycle (Opportunities, Risks, and Mitigation Strategies) Comicboards}}, which illustrated the three key stages of LLMs through scenarios relevant to PRO services \cite{kim2023aligning}. To ensure that workshops were grounded in a basic understanding of LLMs and that participants could meaningfully engage with the concepts, we provided additional scaffolding when needed. In sessions where participants continued to express confusion about what LLMs are after reviewing the comicboards, we briefly displayed and demonstrated a general-purpose LLM (such as ChatGPT), upon request, to offer a concrete reference point. All sessions were conducted remotely via Zoom, lasted approximately 60 minutes, and were facilitated by members of the research team. Sessions were audio-recorded and transcribed for analysis.

\subsection{Participants}
\label{sec:participants}
We employed purposive sampling \cite{palinkas2015purposeful} to recruit 26 participants through CSPNJ, including 16 peer specialists and 10 service users. Participants were recruited to reflect a range of lived experiences and types of engagement with peer support services provided or received within the organization. Consistent with qualitative research standards, recruitment continued until iterative analysis revealed no new codes, themes, or theoretical insights, with the final sample size reflecting this saturation point \cite{caine2016local, rahimi2024saturation}. Given the sensitive nature of the study, all self-reported demographic data are reported in aggregate (see Table~\ref{tab:participant_table}) to protect participant privacy and confidentiality. Peer specialists were recruited via email through existing organizational partnerships, while service users were recruited through peer specialists’ professional networks. Participants received a \$60 gift card to acknowledge their time and contributions; the compensation amount was recommended by organizational contacts based on their experience working with individuals seeking recovery and wellness support. In total, we conducted nine workshop sessions with two to five participants each, along with two individual interview sessions conducted due to participant no-shows and scheduling constraints.

\subsection{Data Analysis}
We employed a reflexive thematic analysis approach \cite{braun2012thematic, braun2019reflecting}. Rather than segmenting analysis by individual comicboards, we analyzed each session transcript in its entirety, following prior HCI work \cite{kuo2023understanding, tang2024ai}. This reflects the fact that participants were, naturally, frequently self-referential across multiple storyboards in their responses within a single discussion. Across 622 minutes of recorded sessions, transcripts were generated using Zoom’s transcription service and reviewed for accuracy prior to analysis. Two researchers then conducted open coding across all transcripts. Higher-level themes were subsequently developed through affinity diagramming, guided by the study’s research questions and emergent relationships among codes. Throughout this process, the research team held regular meetings to iteratively discuss emerging codes and themes. Such collaborative sensemaking is central to reflexive thematic analysis, as it treats analytic differences and diverse perspectives as resources for interpretation through conversation rather than sources of error \cite{braun2019reflecting, mcdonald2019reliability}. Accordingly, we did not calculate inter-rater reliability; instead, codes and themes were refined through discussion and negotiated consensus rather than independent agreement metrics \cite{mcdonald2019reliability}.

\subsection{Positionality Statement}
We acknowledge that our experiences and relative societal privilege shape our research and afford us advantages not shared by our study participants. Our team is trained in the United States across disciplines including Human-Computer Interaction, Computer Science, and Social Work, and includes members with advanced degrees. This study received institutional IRB approval. Given participant vulnerability, we followed best practices from prior research \cite{kuo2023understanding, tang2024ai} and sought guidance from domain experts, including co-authors with social work and occupational therapy training and lived experience as peer specialists. We worked closely with CSPNJ throughout recruitment and study execution, prioritizing a respectful and supportive environment. To build trust, we clearly communicated our role as researchers and emphasized that our goal was to listen to participants’ perspectives rather than evaluate or judge them. To further support comfort and candid discussion, we conducted separate sessions for peer specialists and service users and did not mix these groups within the same workshop.

\section{Findings}

\label{sec:results}

\begin{figure*}
    \centering
    \includegraphics[width=1\linewidth]{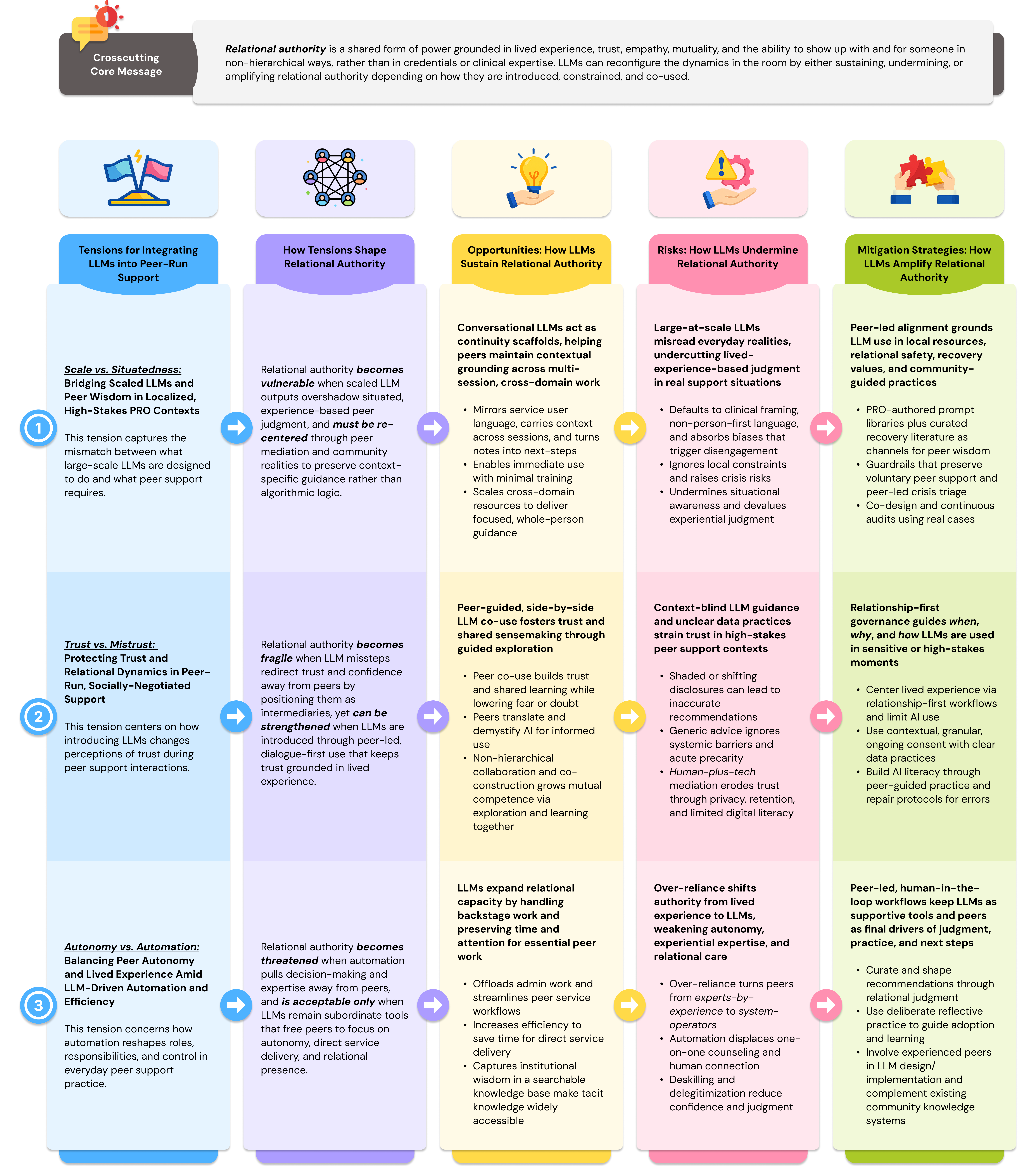}
    \caption{Findings overview illustrating the crosscutting core theme of ``relational authority'' and its manifestation across three tensions shaping LLM integration in peer-run behavioral health services: (1) Scale vs. Situatedness, (2) Trust vs. Mistrust, and (3) Autonomy vs. Automation, alongside participant-identified opportunities, risks, and mitigation strategies that influence whether LLM use sustains, undermines, or amplifies relational authority.}
    \Description{A multi-column diagram summarizing findings on the crosscutting core theme of relational authority in peer-run behavioral health services. The figure is organized into five vertical columns: tensions, how tensions shape relational authority, opportunities, risks, and mitigation strategies. Three horizontal rows correspond to the tensions: Scale vs. Situatedness, Trust vs. Mistrust, and Autonomy vs. Automation. Each row contains text boxes describing how relational authority is influenced, participant-identified opportunities, risks, and peer-led mitigation strategies.}
    \label{fig:findings_table}
\end{figure*}

In this section, we refer to peer specialists as \textit{``PS''} and service users as \textit{``SU''}. Figure~\ref{fig:findings_table} provides a detailed overview of the findings.

Across our findings, one crosscutting core message emerged as \textbf{participants emphasized LLMs can reconfigure in-room dynamics by sustaining, undermining, or amplifying the ``relational authority'' at the heart of peer support, depending on how they are introduced, constrained, and co-used.} Participants described peer support as a form of shared power in which authority is relational and grounded in credibility derived from \textit{``shared responsibility, lived experience, mutuality, and the ability to show up with and for someone in non-hierarchical ways''} (PS13). Unlike authority derived from clinical positions or credentials, participants explained that such \textbf{\textit{relational authority}} is flexible, nonjudgmental, mutually empowering, and growth-oriented. As peers \textit{``lead by lived experience''} (PS11) and \textit{``use that experience to support others in their wellness journey''} (PS13), people may accept guidance when the relationship feels trustworthy and empathetic, built through participatory listening, transparency, integrity, care, and the sharing of expertise in ways that respect the other person’s experience \cite{mead2001peer}. When LLMs enter this space, they do more than redistribute cognitive labor. Instead, they have the potential to influence peer-support dynamics by recognizing the knower in the room, the voice in moments of decision-making, and the presence of real support. Participants saw openings in this reconfiguration: LLMs can scaffold and expand how peers sustain relational work; yet when AI outputs become primary, peer authority can shift from experiential understanding to technical compliance. In PROs, where the relationship itself \textit{is} the intervention, this shift is especially alarming because peer specialists often lack the formal training or institutional authority associated with clinical roles. LLMs risk hollowing out the very fabric that makes lived experience healing, which is personal, contextual, and not universally applicable. Participants therefore called for aligning LLMs with peer language and values to preserve \textit{experiential judgment} as the foundation of relational authority.

This crosscutting theme manifests through three key tensions identified in our analysis: \textbf{\textit{(1) Scale vs. Situatedness}}, \textbf{\textit{(2) Trust vs. Mistrust}}, and \textbf{\textit{(3) Autonomy vs. Automation}}. Together, they trace a throughline from the foundational fit of LLMs in peer-run settings, to trust as a precondition for any shared use, to autonomy as protection for relational work, yielding concrete implications for design and day-to-day practice in PROs. Each tension presents \textbf{\textit{opportunities}} where LLMs sustain peer support or relational authority, \textbf{\textit{risks}} where they create failure modes or relational harm that undermine relational authority, and \textbf{\textit{mitigation strategies}} that reorient practice through stakeholder-grounded design to amplify relational authority. Insights were shaped through the comicboarding workshops, which gave participants sufficient conceptual understanding of LLMs to engage meaningfully with the design space. The visual format helped participants \textit{``visualize and realize what's... going on''} (SU10), allowing many to move from unfamiliarity to workable comprehension even without technical detail. As PS1 explained, \textit{``I don't understand the intricacy of how it does it, I know what it does''}. The comicboards facilitated collaborative sensemaking and helped participants envision real implementation pathways: \textit{``Now I understand a little bit more... how the tool hopes to be implemented''} (PS4). Understanding why the technology might be used also clarified its stakes---\textit{``Just knowing about why it would be implemented... helped me clarify the point of using it''} (PS5)---enabling participants to surface both benefits and harms. These reflections show that the workshops equipped participants to engage critically and constructively, informing the insights presented next.

%\FloatBarrier

\subsection{Bridging scaled LLMs and peer wisdom in localized, high-stakes PRO contexts}

Overall, participants described a core tension in bridging large-scale LLMs with the situated, relational knowledge that defines peer work, emphasizing that peer wisdom is not merely one knowledge source among many but the foundation of relational authority grounded in lived experience, contextual judgment, and an understanding of what is feasible in real circumstances. As conversation starters, LLMs can complement this work by making cross-domain information more accessible, extending relational reach through continuity and memory, and supporting more holistic, present interactions. Yet their centralized models can also displace or distort local knowledge when the outputs reflect clinical or mainstream norms, flatten nuance, and miss the realities that matter in context. Participants worried that this shift risks moving peer support toward algorithmic or clinical problem-solving and away from mutual, experience-based guidance. In response, they described strategies for reclaiming and protecting relational authority, from grounding models in community language and resource realities to ensuring that AI suggestions are always filtered through peer interpretation rather than imposed as generic best practices. These approaches keep decisions shaped through dialogue instead of algorithmic instruction, preserving relational authority in community expertise. Across these accounts, the goal was not to place authority in the model but to re-center it in the relationship, using LLMs only insofar as they help make peer wisdom more coherent, actionable, and contextually grounded.

\subsubsection{Opportunities: Conversational LLMs Strengthen Peer Relationships and Support Holistic Cross-Domain Guidance}
\textbf{Participants valued LLMs' semantic and conversational capacities when they reinforced the relational foundation by mirroring service users’ language, simplifying dense information, and assembling actionable summaries from loose notes in ways that complemented, rather than redirected, the peer's approach.} As one PS explained, \textit{``Being more conversational is way more beneficial because peer work is all about mutuality and meeting people where they're at''} (PS11). LLMs tolerated ambiguity, carried context across interactions, and adapted tone to match the user, which several service users said made materials easier to understand. SU7 emphasized synthesis, \textit{``The AI detects the strong key points within the conversation... what that person needs... and their goals.''} These outputs also served as openings for further dialogue, enabling peers to ask more attuned questions and continue building support through conversations. PS2 highlighted continuity, \textit{``As you use the AI more, it gathers all the information and piggybacks on the last thing that you gave it.''} Tone also mattered for comfort in peer work, since SU1 felt the words were \textit{``still genuine''} yet delivered \textit{``in a soft tone and touch way,''} and that language choice \textit{``can change the feel''} of the interaction. Paired with peer judgment, an hour of scattered updates became a brief summary for the service user, a checklist to follow through, and a living case record, helping peers grasp what mattered to service users while still maintaining interpretive control.

\textbf{LLMs give under-resourced PROs a natural-language interface that enables immediate use without specialized training while also scaling access to a broad, cross-domain knowledge base that individual sites could not compile on their own, helping peers sustain breadth of expertise and draw on expert-curated resources when needed}. \textit{``An LLM would make things centralized... with the eight dimensions of wellness, we need something that can combine them all''} (PS11). Participants described how this breadth matters because peers provide wide-ranging support, and LLMs can help identify options that fit a person's lived circumstances while guiding meaning making in ways that reinforce experiential judgment. Resources feel \textit{``at our fingertips,''} and that looking at a person as a whole allows peers to \textit{``find skills and strategies to address and locate additional services, including housing, financial, and whatever it is that they may need''} (PS7). When LLMs organize information across life domains, they help peers understand the whole situation rather than treating support as isolated problems. Tools that \textit{``help identify what realistic goals we can help in the different dimensions''} support readiness planning (PS2). Compared with the open web, \textit{``Google search gives you too many choices''} (SU4), a helpful system would let users type a need and see a focused list \textit{``that pertains to all of that... you can recertify, you could do this and you could do that''} (SU2). It can assist with concrete tasks, \textit{``practice interview questions, very specific things''} (PS6), and prompt the right questions that build motivation in context, for example during a hospital stay when the near-term goal is food access after discharge (PS5).

\subsubsection{Risks: Large-at-Scale LLMs Misread Everyday Realities and Undercut Peer-Support}

\textbf{General-purpose LLMs are trained on mainstream internet data and broad benchmarks, not culturally competent, trauma-informed, recovery-oriented peer support, and participants observed that models often default to clinical terms like patient, compliance, and treatment rather than peer language like service user, choice, and recovery journey.} PS11 said, \textit{``Peer work is not at the forefront... when people think of mental health, they think patient and medicine... we are not clinicians... if people struggle to get that, I can see how LLMs would struggle as well.''} LLMs can impose a medical-model frame that peers reject, shifting power toward the model by implying clinical knowledge as more authoritative, turning mutuality into hierarchy, and undermining relational authority with diagnostic expertise. Language choices also matter for emotional safety. PS2 said, \textit{``Instead of saying someone is paraplegic, we say a person experiencing paraplegic behavior or incapacities... incorporate that language.''} PS11 added, \textit{``We use person-first or person-led language so people are not defined by their diagnosis.''} When LLMs overwrite these norms, the interaction can lose the cultural grounding that makes peer support safe. Moreover, peers warned that models can absorb biased inputs: \textit{``There could be biases of the individual inputting the information... maybe even watch the language that you use''} (PS8). The misalignment in the culture and language can cause harm and disengagement. PS11 cautioned that terms like suicidal ideation can be \textit{``triggering''} and that people will disengage \textit{``if they feel it is not meeting them where they are.''} Service users added that LLMs can read lived realities as pathology and flatten context by \textit{``collecting bits and pieces and putting you in a category''} (SU7). Without domain data and community wisdom, models risk misinterpreting trauma responses or survival strategies as symptoms, stripping these behaviors of meaning and reinforcing stigma.

\textbf{When models flatten nuance or disregard real-life constraints that peers consider in grounded support, they undermine situational awareness and devalue experiential judgment by producing advice that sounds confident and authoritative but is impossible to act on.} LLMs can deliver the same outputs unless heavily specified and wellness areas are \textit{``so broad''} with \textit{``subcategories in each dimension''} (PS2). Service users echoed sameness and oversimplification, saying AI \textit{``responds in the same manner''} and gives \textit{``a straight yes or no... but in reality, there's a lot of gray lines''} (SU4). Participants illustrated how models routinely overlook systemic barriers, such as unaffordable services, discriminatory environments, and unsafe or nonexistent family support, recommending services \textit{``a town or a city over''} with \textit{``no bus routes''} (PS12), or assuming phones and stable support \textit{``LLMs just focus on the middle class rather than people who have very limited resources... sometimes they don't even have the phone to call''} (PS11). This access gap is exacerbated because many service users \textit{``can't even speak precisely for the LLM to understand everything''} (SU4) and have low digital literacy \textit{``I've had guests that I've had to teach how to use Google''} (PS2). When guidance drifts from reality, the gap between algorithmic logic and lived experience widens, weakening relational authority as peers must mediate or reinterpret outputs that already misrepresent actual needs and correcting a tool that appears more knowledgeable than they are. The risk intensifies because participants were particularly cautious about misdiagnoses that could surface the wrong supports: \textit{``if somebody got misdiagnosed, then it's using that false diagnosis as the most important thing to find resources''} and stressed that systems should not default to clinical pathways without request (PS8). Such behavior collapses boundaries that protect peer support from clinical authority, and these shortcomings are most concerning during crises, where peers emphasize trauma-informed support: \textit{``Instead of treating people, it's more about how to live a manageable life from a peer lens''} (PS11).

\subsubsection{Mitigation Strategies: Peer-Led Alignment Grounds LLMs in Local Resources, Relational Safety, and Community-Guided Practice}

\textbf{Participants envisioned a peer-lens LLM that uses prompt engineering, fine-tuning, and retrieval-augmented generation (RAG) to adapt general resources to local PRO needs so the model learns from peers rather than positioning itself above them, becoming a channel for peer wisdom while keeping peers as the model for what good support looks like.} PS14 urged person-specific prompting and peer-lens system instructions because \textit{``getting to know this person is the most fundamental thing''} and doing so \textit{``might give you a better nuanced response more tailored to the individual's context.''} They described prompting as a form of peer reasoning, \textit{``sorting it out in one’s head to figure out what they need,''} making prompt engineering an extension of peer judgment that keeps model outputs tied to peer framing and recommendations contextualized through peer interpretation. PS2 emphasized learning from lived experience and called for grounding outputs in what people are facing \textit{``at this moment.''} To shift from \textit{developer-led fine-tuning} to \textit{community-centered alignment}, peers proposed PRO-specific prompt libraries authored by experienced peer specialists with step-by-step, recovery-oriented practices, paired with curated recovery literature and authentic peer-support conversations. Examples included harm reduction beyond substances (PS1), tapping techniques, CBT and DBT, and mindfulness and meditation (PS2). Using peer recovery language \textit{``really does make a difference on how people respond''} (PS2), conveys \textit{``possibility for change''} and \textit{``hope''} (PS3), and person-first phrasing helps \textit{``decrease that stigma language''} (PS1). Beyond alignment, participants emphasized making guidance actionable and kept up to date, backed by organization specific resource libraries. PROs already do this, \textit{``trying to keep the resource library as up to date and as informational as possible''} (PS5). Governance matters \textit{``managers go in to update that information on a regular basis,''} with \textit{``a protocol,''} \textit{``weekly reports,''} and \textit{``a user identity on whoever entered it''} to ensure local accuracy (PS11). Peers asked for inventories of open ended wellness questions and tracking \textit{``what resources or strategies they've already used, or tried and failed with''} to \textit{``narrow down what this person would need''} (PS2). 

\textbf{With increased capability comes a need for clear guardrails that preserve the voluntary, relational nature of peer support and avoid clinical overreach.} As PS11 observed, liability concerns can lead some chat systems to \textit{``automatically call the police because they don’t want to be held responsible for anything that could go wrong,''} yet in sensitive situations that response can endanger the person seeking help and do more harm than good. They pointed to mobile crisis outreach models that send a peer with a clinician and a psychiatrist to \textit{``avoid police intervention.''} In acute risk moments, a peer-informed triage should lead: \textit{``the first thing you always ask is are you safe''} and responses should first validate and gauge whether someone is active or passive rather than rushing to problem solving (PS11). Peers reiterated that the relationship must remain primary, \textit{``build a social connection and help guide them across multiple dimensions of wellness''} (PS13), and LLMs should integrate with that human work, \textit{``there are certain things that you need human interaction for but the computer might be able to learn''} (PS2).

\textbf{Lastly, participants insisted on \textit{lived-experience-in-the-loop} oversight, partnering with PRO communities to co-design, test, and regularly audit LLMs with both peer specialists and service users before and after deployment, so feedback continuously refines responses and surfaces community- and identity-specific impacts.} PS11 said, \textit{``Nothing about us without us... we need to be involved in the testing and training from the beginning,''} including teaching language boundaries \textit{``we never say patients... that could be the hard line.''} PS2 added that people \textit{``actively in the field of mental health''} provide an \textit{``eagle-eye view''} to keep recommendations aligned and \textit{``actually feasible for the service users.''} PS10 noted the access layer beyond an address \textit{``how do I get there... access is so important.''} To keep alignment active, PS12 suggested logging real questions and barriers as \textit{``additional material''} that can be shared to the technical team, so iterative audits and updates continually incorporate community voice and lived practice.

\subsection{Protecting trust and relational dynamics in peer-run, socially-negotiated support }

Across sessions, participants emphasized trust as the core foundation of peer-run support, where the relationship is the intervention rather than simply a medium. Recovery happens through authentic connection, empathy, and shared experience, emerging not only from what is said or done but from being seen, trusted, and accompanied by someone who has been there, a relational act that carries mutual recognition, hope, and solidarity. Introducing LLMs during co-construction can reinforce that trust when they operate in alignment with the peer to extend relational work, with the model being trusted \textit{through} the peer, \textit{with} the peer, and \textit{in service of} the peer-service user relationship. At the same time, participants noted that LLM errors and irrelevant or inappropriate responses can weaken peer specialists’ credibility because peers become the face of the mistake. When generic advice or misplaced certainty makes service users feel unseen, trust can shift away from peers as their lived expertise is undermined by harmful outputs, positioning them as intermediaries for algorithmic guidance rather than trusted partners, which compromises relational authority. Participants therefore stressed dialogue-first workflows that keep human judgment primary, protect service users’ agency through voluntary, mutual engagement with LLMs, and treat breakdowns as opportunities for community involvement, reflective practice, and relational repair. Together, these perspectives highlight that protecting relational authority requires keeping trust and lived experience as the central epistemic grounding so that LLMs function as relational amplifiers rather than substitutes.

\subsubsection{Opportunities: Peer-Guided Co-Use of LLMs Builds Trust, Shared Learning, and Mutual Competence}
    
\textbf{Introduced through peer relationships and modeled use, LLMs can strengthen trust by co-constructing support and transferring confidence from peer to tool as the technology becomes trusted through the peer.} PS11 captured this dynamic: \textit{``Working with LLM would be the same as teaching a skill... showing them first and modeling it builds that safety and trust so they feel like they can do it on their own.''} Trust rises with side by side use, as peers can act as trusted guides who demystify capabilities and limits, \textit{``educate around the myth in people’s minds about AI,''} and introduce LLMs in supportive, non-judgmental ways (PS13). As participants explained, \textit{``human interaction is the advantage... peer specialists can really be the translator for the person... help them to be educated consumers of AI.''} Keeping the person involved is also a safeguard that reduces fear and deepens connection. \textit{``You're building that relationship at that moment... by being involved in it, that reduces some of that fear that someone may encounter with it''} (PS14). In this dynamic, the LLM becomes a partner in relational work while the relationship remains the container for trust, providing reassurance and affirmation that protect relational authority.

\textbf{With mediated exposure, successful navigation, and visible safeguards, comfort can grow, fear can ease, and service users may adopt a self-efficacy mindset such as \textit{if someone like me can do this, maybe I can too} while trust in the peer remains central, creating a co-constructed space where people can explore AI, voice fears, and learn safely.} PROs’ nonjudgmental, strengths-focused culture lowers barriers to engagement, \textit{``we welcome everyone... people here are very accepting and tolerant''} (PS15). Service users often mapped LLMs’ conversational flow onto this familiar relationship building, while peers noted LLMs helped clarify next steps and warm the wording, which can increase perceived credibility \textit{``help peer specialists feel more credible''} (PS3), refresh basics \textit{``remind us of the eight dimensions of wellness''}, and render guidance in \textit{``plain language... warmer and more empathetic''} (PS11). When peer specialists explore LLM capabilities with service users, the co-engagement creates collaborative learning and reflective practices that strengthen their relationship. As PS1 noted, \textit{``One of the opportunities that I see is the opportunity for collaboration between the peer support worker and the person served.''} This collaboration is explicitly non-hierarchical: \textit{``There's no power differential... me using the tool and then showing them... both [learning] how to use it [and seeing] the benefits definitely will build that collaboration and understanding''} (PS11). This amplifies relational authority because the LLM becomes a medium for relational closeness, with peers walking alongside rather than instructing from above. Together, open co-use builds shared understanding, increases trust, and develops mutual technological competence.

\subsubsection{Risks: Context-Blind Guidance and Unclear Data Practices of LLMs Strain Trust That Peer Support Depends On}

\textbf{PROs are uniquely high-stakes because eligibility and care hinge on sensitive details, such as income, legal issues, criminal justice involvement history, and bank balances, which many people understandably withhold or shade, making accurate disclosure difficult and consequential and potentially compromising screenings and downstream recommendations.} PS15 explained, \textit{``Trying to get accurate information out of people is hard... with their legal issues... if they have any felonies or whatnot demographics... how honest are people gonna be... they're not gonna tell me that.''} When peer specialists encounter embellished or incomplete narratives from service users, LLMs can produce confident but wrong recommendations. PS5 observed, \textit{``Maybe they accidentally said something that is contradictory about their background... may be confusing when it comes to figuring out benefit eligibility.''} People’s shifting situations and preferences can also create apparent contradictions that LLMs may misread as inconsistencies rather than evolving context. In these moments, LLMs risk substituting relational understanding with algorithmic literalism, missing the tacit cues peers use to interpret what is said and unsaid, especially in high-stress conversations where people blend literal events, metaphor, and experiences that are hard to verify and difficult even for humans to parse: \textit{``They're like talking about certain situations that may not always make sense... you have to believe them and follow the steps regardless... I don't know if that could interfere with the AI''} (PS8). This undermines relational authority because the tool cannot honor the nuance peers rely on to make sense of complex or emotionally charged disclosures.

\textbf{Access varies widely across places, making uniform LLM suggestions brittle because service users face different resource landscapes and many live in acute precarity. When algorithmic outputs appear more expert despite failing to reflect real conditions, one-size-fits-all recommendations can push people toward impersonal advice and undercut peers' situated knowledge.} When LLMs give suggestions such as \textit{``hey, are you focusing on your physical health,''} this might trigger people when they \textit{``aren't sure if they are gonna eat tomorrow''} (PS2), many \textit{``can’t go to a yoga studio or afford a gym membership,''} and generic wellness advice like \textit{``get 8 hours of sleep''} simply isn’t feasible \textit{``if someone is worried about where they're gonna sleep that night''} (PS13). Such errors can sever hard-won trust and discourage future help-seeking, compounding harm by neglecting local constraints. 

\textbf{Technical misalignments in LLM-mediated guidance reverberate through the relationship by making service users question whether they’re receiving authentic peer support or AI-filtered advice, shifting bilateral trust from \textit{human-to-human} to \textit{human-plus-technology}, and compromising relational authority as people interpret the peer's attention as divided or replaced by machines.} PS9 voiced the concern: \textit{``My fear is that a peer provider sits in front of somebody... typing something on the internet [of] what we think this guy may need, instead of building that relationship and understanding what this guy tells me he needs.''} Participants described risks when model certainty misleads users. LLMs' authoritative tone can cause service users trust the system over the peers or pressure peers to defer even against lived knowledge. This reverses relational hierarchy and diminishes lived experience's role, positioning the model as expert and peers as subordinate interpreters. Because service users initially extend trust to LLMs via trusted peers, harmful recommendations can damage confidence in both the person and AI, \textit{``if the LLM gives some ridiculously broad or hallucinated response, it will a hundred percent lessen the trust between the peer provider and the people that they’re trying to serve''} (PS11). This creates a \textit{``shoot-the-messenger''} dynamic: \textit{``It could potentially lead to the mistrust being targeted towards the person translating''} (P11). Relational authority rests on the belief that the peer understands one’s situation and offers guidance grounded in that understanding. When LLM failures are misattributed to peers lack institutional credentials, service users may associate the harm with the peer, making them appear unreliable or uninformed even when the failure originates from the model. In this unique setting, the risk is also asymmetrical as peer specialists operate the technology, but the service users are the real ones who bear the direct harm. 

\textbf{Participants raised acute confidentiality risks, including data retention and unauthorized access, which feel especially threatening in communities with histories of institutional harm. These fears can extend to peer specialists, who may seem aligned with systems that monitor or mishandle sensitive details, putting relational authority at risk because peers are expected to protect service users from such systems, not feed information into them.} Given the sensitive nature of information stored in PRO contexts, service users frequently must exchange details about their immigration status (often undocumented), justice involvement records, and medical histories, and they question whether systems truly honor deletion and consent as they wish (PS2, SU5, SU10). Concerns extend to accidental exposure of personal recovery histories: \textit{``Someone out there has a computer knowledge could hack probably anything that I can't even imagine because it would be all your information in one place''} (PS2). Layered on top of this are community-level anxiety and fears: \textit{``This is a new tool, just hearing that can just kind of conjure up different thoughts in people's minds''} (PS3); \textit{``We have a lot of members... who won't even use a cell phone because of paranoia''} (PS8); and \textit{``Tech literacy is pretty low... even just making an email account is a struggle for some people''} (PS11). At the same time, data ambiguity and fast-changing resource rules make AI fragile: \textit{``[LLMs] can only provide what’s been given... what if they think that someone’s eligible and then they’re not? Because they don’t have clear rules from these places to begin with''} (PS11). In tight-knit PRO communities, a single failure can spread and deepen mistrust. Even small breakdowns can erode trust in both peers and AI, and harden opposition to adoption.

\subsubsection{Mitigation Strategies: Ensure AI Serves the Relationship by Centering Lived Experience, User Agency, and Relational Repair}

\textbf{Overall, participants emphasized centering peer specialists’ lived experience before any AI input, proposing a \textit{story-first, relationship-first, LLMs-later workflow} that keeps mutual attunement at the heart of encounter to protect relational authority.} They highlighted \textit{``lived experience is the differentiator between care work and clinical work''} (PS11). To further prioritize experiential knowledge, participants advocated designing the workflow so peer specialists translate their own insights into actionable guidance rather than translating LLM outputs. In practice, this can begin with PS2’s first-meeting script: \textit{``The first things I ask is, ‘What would an ideal day look like for you?’''} because \textit{``whatever comes first to mind usually aligns with what they believe in... and they often already know their barriers and can tell you.''} Participants also recommended limiting AI consultation to between-session preparation and follow-up to protect direct \textit{peer-to-peer} time, ensuring that AI must preserve and not invade relational presence (PS11). 

\textbf{Instead of standardized policies, participants framed consent as contextual, granular, and ongoing, urging that service users control \textit{if}, \textit{when}, and \textit{how} LLMs are used—including opting out entirely, limiting use to specific applications, choosing assistance levels, and even entering their own information—so engagement remains voluntary, preserves agency, avoids HIPAA concerns, and reinforces the session as collaborative relationship-building rather than something coerced or outsourced to a machine.} To operationalize this, systems should clearly explain storage, access, confidentiality, and retention; obtain informed, per-type consent that acknowledges recording and deletion policies. SU3 tied comfort to time limits: \textit{``If it’s just stored for a short period, I could be fine... over a year would be too long.''} Access should also be narrow, SU6 emphasized: \textit{``Your information is shared with that center or that peer specialist only, and not other peer specialists.''} People also wanted a clear choice: \textit{``So maybe giving people that option, like this version will not memorize, but it will not be as therapeutic or as effective as the version that is learning you, cause it is a choice that you have to make''} (PS2). To reduce identifiability, PS6 suggested: \textit{``We can use initials to call this person... your actual name is not associated with your information,''} PS2 added that ranges beat specifics: \textit{``Sometimes people feel more comfortable if it's giving a range rather than specifics''}, and PS3 endorsed numbered identifiers: \textit{``Just accessing number one to pull up the information so no one knows who that is.''}

\textbf{Critical AI literacy strengthens the peer’s role as guide and interpreter, centering relational understanding as the basis for deciding what matters. When LLM errors become moments for transparency, reconnection, and repair, peers demonstrate accountability and care, making clear that the peer remains the responsible and trustworthy partner.} Participants stressed that hands-on practice with LLMs raises overall AI literacy and reduces fear: \textit{``I see it already happening... when we first said AI... many said what no no... but as we manualize it we see more and more people... learn [to use LLMs] in an educated way... not throwing everything in it... to reduce that irrational paranoia and fear''} (PS13). They also recommended framing AI as a collaborator rather than an authority: \textit{``Integrated technology into your workflow, but in a way it actually combines your own strengths with the technology strengths''} (PS11). When situations are too high-stakes or complex for LLM assistance, participants emphasized having \textit{failure-response procedures---immediate acknowledgment, clear explanation, co-created correction steps, and relationship repair}---so AI failures become opportunities for trust-building rather than rupture (PS11). Correction is co-owned with the service user: \textit{``‘what do you think about that?’ and using what [we] learned from past interactions''} (PS3). Taken together, these practices turn AI failure into a structured moment for relational repair. 

\subsection{Balancing peer autonomy and lived experience amid LLM-driven automation and efficiency}

Where the first two tensions examine knowledge fit and trust, this tension focuses on how automation redistributes authority in everyday practice. Participants described success in peer sessions as preserved autonomy, respect, and voice rather than time saved. They emphasized that automation is acceptable only when peers retain control of the workflow, decision making remains with peer specialists, and processes keep humans in the loop so that boundaries for care and accountability stay intact. They saw value in LLMs when automation lifted routine tasks off peers, allowing more focus on human connection by returning time and attention to peer specialists’ relational work rather than the system’s technical work. Relational authority also grows when LLMs serve as backstage tools that capture tacit peer wisdom, reinforce lived experience as shared knowledge, and help peers show deeper understanding of service users’ situations during support. On the other hand, participants cautioned that over-reliance on LLMs risks shifting relational authority from lived experience to the system. Automation can threaten the dignity-based and autonomy-preserving ethics of peer work, since deskilling and diminished confidence weaken relational credibility, and training new peers primarily as LLM operators undermines the long-term transmission of the experiential expertise that defines peer identity. Participants therefore emphasized keeping LLMs subordinate to, shaped by, and in service of the relational authority that defines peer support by designing transparent human-in-the-loop workflows that hold peers accountable as the agents of meaning making and treat LLMs as practice-sharpening tools rather than sources of authoritative instruction. Overall, the central aim is to maintain clear role boundaries between human judgment and automated assistance and to support an ethical distribution of responsibility so that LLM integration reaffirms peer identity and credibility while amplifying peer interpretation and relational connection.

\subsubsection{Opportunities: LLMs Expand Relational Capacity by Streamlining Workflows and Scaling Institutional Knowledge}

\textbf{At PROs, LLMs help bridge service gaps created by high caseloads and limited resources through synthesizing large volumes of information, generating actionable suggestions, and streamlining administrative tasks, freeing peer specialists to focus on direct service delivery, serve more service users efficiently, and invest more in relational authority.} Peers also want structure so personalization improves: \textit{``We're identifying different barriers like transportation, finances, hospitalization, incarceration... if we could set the who, what, when, where, and why, it helps providers know what questions they should ask and helps the AI answer in the best way''} (PS7). Participants also mentioned that LLMs can help sustain engagement over time by automating scheduling and reminders, prioritizing follow-ups by urgency, and using participation data to flag timely outreach (P2, P13). Throughout this process, LLMs should carry context across sessions. PS2 explained, \textit{``All these different areas of your life affect one another... if you’re not doing good in your environment or occupational well-being, it usually flows into your financial, emotional, and social well-being... [LLMs] could help people review in a session... with small, plain-language tidbits and real-life examples... it would guide practice''} (PS2). Participants reported that LLMs can accelerate the routine but critical paperwork in peer services. They can listen to sessions and draft recovery-oriented notes in peer language, then repurpose the content for other formats. From conversational notes, LLMs can prefill intakes, assessments, and service plans, flag inconsistencies for review, and push structured fields directly into records, saving staff time (PS9). PS2 imagined an assistant that can \textit{``scan all the notes... [to] get a full-spectrum picture of this person... even prior to meeting.''} Some service users see LLMs as a neutral backup when a peer needs a moment, since the AI \textit{``doesn’t have feelings''} and simply returns an answer (SU4).

\textbf{LLMs maintain a searchable knowledge base that captures institutional wisdom and aggregated patterns, such as what supports justice-involved users in employment, making tacit knowledge accessible without manual documentation, broadening expertise across communities in ways that democratize access and uphold peer-run values, and protecting rather than displacing relational authority.} By analyzing these patterns, LLMs can flag systemic barriers and suggest when and how to reach out most effectively; as PS8 described, by \textit{``input[ting] a past scenario and compar[ing] it to the path actually taken.''} This helps people move from patterns to practice: \textit{``The peer specialists can then say, ‘This is what the LLM shows, so here is how we can approach transportation or legal court case barriers’''} (SU2). People also see value in speeding up bureaucracy as LLMs could help \textit{``find a grant or an agency that would help me financially [or] emotionally''} (SU4). During challenging moments, peers can also quickly query evidence-based techniques, de-escalation strategies, or policies in real time without interrupting service. For complex benefits like Medicare and Medicaid, LLMs can clarify eligibility and options for what exactly is appropriate for a service user (PS9), provide step-by-step instructions such as what IDs to bring to which office addresses (SU6), and respond quickly when agencies are unreachable: \textit{``I can’t get anyone on the Social Security phone... [LLMs] can help me figure out what I need to do without speaking to a live representative''} (SU4). Lastly, LLMs can turn anonymized cases into role-plays, discussion prompts, and training materials, support employment tasks such as interview practice, applications, and resume tailoring, and quickly brainstorm situation-specific options for needs like high anxiety (PS6, SU7, PS7).

\subsubsection{Risks: Over-Reliance Shifts Authority from Lived Experience to LLMs, Weakening Autonomy, Expertise, and Relational Care}

\textbf{Participants cautioned that over-reliance on LLMs can shift the relational foundation from \textit{peer-as-expert-by-experience} to \textit{peer-as-operator}, directly eroding relational authority as peers move from drawing on lived experience to merely replaying model outputs and becoming intermediaries between technology and service users.} As PS2 warned, \textit{``The harm [is] if it’s being used where you’re no longer using it, and it’s just doing it for you.''} PS11 emphasized that success must be measured by whether people feel respected and self-determined, not by efficiency gains: \textit{``If [the LLMs] cost dignity... that's when it becomes a failure... more than how much time it saves, the success should be measured by whether people feel respected, that their autonomy isn't being completely run over... [they should] have the final word... and [the LLMs] should just be assisting, not controlling.''} Participants added that automating documentation, resource navigation, and scheduling can erode peer specialists’ confidence in their own expertise, critical thinking, and professional judgment, thereby diminishing one-on-one counseling: \textit{``Too much [reliance] on the technology alone... you would just keep referring back to this instead of the one-on-one counseling''} (SU4). Concerns extended to dehumanization and deskilling, as PS1 noted \textit{``the loss of human interaction could be one of the biggest risks... if humans don’t get their stuff together with technology, technology will be replacing us as well''}; \textit{``Basically, it's like you're just depending on the computer to do the work and not the individual''} (SU5). Several also warned that training new peer specialists primarily as LLM operators could undermine the inheritance of deep experiential expertise that relational authority depends on, since LLMs cannot supply the shared peer knowledge and lived experience that define peer support.

\subsubsection{Mitigation Strategies: Peer-Led, Human-in-the-Loop Workflows Keep LLMs as Supportive Tools Rather Than Drivers of Peer Practice}

\textbf{Participants recommended framing LLMs as tools that amplify, not replace, peer specialists’ existing expertise, creating a model where peers curate and shape recommendations through relational judgment rather than execute automated instructions, keeping them firmly in the driver’s seat.} PS9 urged a person-centered approach in which prompts \textit{``remind the peer to ask more personal questions''} while PS2 emphasized \textit{``it would be important to still... use their own knowledge... to make sure it’s individualized and that we are guiding it through.''} They also want to design workflows where peers actively approve or modify suggestions, rephrasing in their own words or supplementing expertise with AI outputs rather than passively implementing them. PS7 echoed this, \textit{``[Service users] wanted it to be from you... genuinely from your words [not] robotic answers... even though you’re taking words and concepts... remember to add personal spins to everything.''} PS11 advised contextualizing LLMs, \textit{``Give it the exact steps of what to do... people overestimate AI's capability... I really try to just break it down like how I would when teaching a toddler... with the supportive housing, they were giving me examples that just wouldn't work with the population... so I told the AI, these people are going to be cynical... jaded... burnt out, they're not going to want to trust anything... once I gave that sort of reconfiguration, they now know how they should respond.''} 

\textbf{Participants advocated for active sensemaking use that enhances relational authority by treating LLM interactions as \textit{deliberate learning moments} that sharpen practice: analyzing why a recommendation works, noticing when it conflicts with lived knowledge, and building expertise through structured reflection.} PS12 suggested viewing an LLM’s output as a hypothesis, validating it against the service user’s context and cross-check sources for iterations. They reflected on a previous experience: \textit{``It gave me two different URLs for the website of a housing authority and neither is correct... then an address that I don't think exists after Googling... and finally a phone number that I dialed and it’s inactive.''} To support continuous improvement, PS14 proposed that PROs periodically draft and test new resource entries to identify additions for the model and flag items that are missing or no longer available. To reduce over-reliance, PS2 emphasized guided adoption: \textit{``We could avoid that risk... by guiding, and having setups where the peer specialist is learning how to use it through a guiding method rather than just only relying on it.''} Moreover, PS11 outlined a consent-driven flow for in-the-moment support: \textit{``Start by gauging the concern or major issue in that moment''}; after identifying the main issue, ask what outcome they want from the help: \textit{``Do you want to cancel the appointment or go to the doctor?''} Then work it out together and get approval at each step: \textit{``How does this feel to you? Is this a direction that we should be going in?''} 

\textbf{Participants urged involving experienced peer workers in LLM design and implementation---from model training, iterative audits, to updating resource banks---so adoption aligns with peer-support values and preserves professional autonomy. They stressed learning resources directly from community members, not only from AI, because PROs rely on word-of-mouth, lived knowledge networks, and embodied experience for reliable information, and LLMs should only complement these ecological knowledge systems to protect situated and relational epistemology of peer support.} PS8 said, \textit{``The easiest way to learn about resources is from the people using the resources... unless you make that phone call and hear directly from the person giving services, you’re not going to get an accurate understanding of the program or the resource.''} In addition, peer specialists emphasized that LLMs should handle only routine tasks to free them for higher-value experiential work, rather than automating core peer-support functions, suggesting AI’s potential to protect experiential practice. PS13 explained, \textit{``We could definitely help create a workflow to be programmed into this... to help bring up prompts and processes that help peer supporters be more faithful to their role... the resource is important, but I think more important is creating the relationship.''} Lastly, participants called for future LLM design, development, and deployment to align with recovery principles of choice, hope, and self-determination, supporting peer specialists in empowering service users without heavy technological dependency: \textit{``We always say... it’s nothing about us without us... [peers] should be a part of it from the beginning... it should be peer-centered and make sure those principles are aligned with what peer-work principles are''} (PS11). Following these principles, LLMs can amplify relational work and preserve peers as the holders of ethical relational space.
\section{Discussion}
\label{sec:discussion}

Across our findings, \textit{\textbf{relational authority}} emerged as the central mechanism through which peer support operates and through which LLMs can either sustain or unsettle care (see Section \ref{sec:results}). Unlike clinical authority, it is enacted through how peers listen, translate, reassure, and make sense of complex situations with service users, and it is preserved only when peers remain the primary interpreters of meaning, context, and care \cite{mead2001peer}. Introducing LLMs into PRO settings therefore does more than add new capabilities. It reconfigures how expertise is signaled, how trust circulates, and how decisions are co-authored in practice. In the following sections, we build on this theme to examine how LLMs’ conversational performances intersect with peer relational labor, how trust is co-constructed or strained within a \textit{human-plus-technology} triad, and how centering lived experience in design can protect peer autonomy as new forms of automation enter everyday work. Together, these analyses illuminate how LLM integration can reshape the relational authority structures at the core of peer-run behavioral health support.

\subsection{From Model Capabilities Development to Collaborative Integration}

Our findings reveal that, for both peer specialists and service users, LLMs differ qualitatively from earlier task-specific or rules-based AI systems. Most saliently, LLMs combine semantic reasoning with turn-by-turn conversationality. Participants described how this pairing allows LLMs to \textit{perform} aspects of relational labor, such as tracking prior turns, mirroring language, and offering stepwise next actions \cite{bickmore2005establishing}. These affordances can make LLMs feel approachable, and at times even caring. In under-resourced PROs, participants noted that such features could help scaffold documentation, organize casework, and support peer specialists in preparing for or reflecting on sessions. 

At the same time, participants emphasized that these relational performances remain surface-level. While carefully constrained LLM outputs can scaffold conversations or help draft supportive language, they are fluent yet fundamentally non-relational. LLMs lack the lived experience, accountability, and mutual understanding that define peer support, and their apparent empathy risks blurring boundaries when peer specialists or service users over-ascribe insight or care to the system. In PRO settings---where trust, voluntariness, and lived experience are not ancillary but constitutive of practice---this distinction is critical.

Because relational authority, rather than model capability, drives care in PROs, integration must ensure that model outputs never displace peers’ \textit{\textbf{experiential judgment}} or the mutuality that makes the work meaningful. When service users attribute empathy or intent to an LLM, they risk mistaking fluency for care \cite{jakesch2023co}. When polished outputs overshadow a peer’s own voice, they can diminish the value of lived experience. These risks are not merely cosmetic. In PROs, the relationship \textit{is} the intervention. If integration is not relationally grounded, LLMs can silently re-center algorithmic or clinical authority, shifting who defines support in the moment and weakening the cultural nuance, shared experience, and trust that constitute peer expertise. Undermining this authenticity can erode trust and compromise care \cite{bickmore2005establishing, chen2022trauma}.

Participants articulated these breakdowns through three distinct LLM failure modes:
\begin{enumerate}
    \item \textbf{\textit{Inaccurate.}} Fabricated or misleading outputs can cause material harm in high-stakes contexts such as housing, benefits, or legal aid, where errors are often difficult to detect or correct mid-interaction \cite{vasconcelos2023explanations}.
    \item \textbf{\textit{Irrelevant.}} Generic, clinical, or oversimplified responses can miss the nuance of trauma-informed, community-grounded, recovery-oriented support and feel dismissive or alienating \cite{chen2022trauma}.
    \item \textbf{\textit{Inappropriate.}} Context-blind responses in crisis situations (e.g., suicidal ideation, substance use, domestic violence), such as defaulting to police involvement, can retraumatize service users and escalate harm, particularly for individuals from overpoliced or marginalized communities \cite{chen2022trauma}.
\end{enumerate}

These are not simply usability bugs; they are relational ruptures that risk reintroducing the very institutional harms peer support was created to counter. They also represent moments where model outputs try\textit{---implicitly---}to speak with an authority that is not theirs.

We therefore argue that what is needed is not only improvement in model capabilities, but \textit{\textbf{collaborative integration}}. Here, integration is an infrastructural question about preserving the peer’s relational authority so LLMs assemble, translate, or prompt without ever defining the logic of support on their own. Participants envisioned LLMs as generative tools that help organize information or surface possibilities, with outputs always filtered through peer judgment and situated in lived experience. This vision repositions LLMs not as clinical decision-makers or neutral recommenders, but as conversational collaborators that remain subordinate to peer interpretation. Such positioning protects peer autonomy, preserves relational integrity grounded in experiential expertise, and leverages AI to expand capacity without displacing the care or authority that reside in the peer relationship.

\subsection {Trust is a Relationship, Not an (LLM) Feature}

Introducing LLMs into peer-run behavioral health services does more than add a new tool. It reshapes the social mechanics through which care is delivered and received. In PRO contexts, trust is not a static attribute or product feature. It is \textit{\textbf{co-constructed}}, \textit{\textbf{continuously negotiated}}, and \textit{\textbf{grounded in shared lived experience}}, and it is the mechanism through which relational authority is enacted. Trust signals who is believed, who defines meaning in the moment, and whose interpretation guides next steps. Bringing LLMs into this space transforms a dyadic \textit{human-to-human} relationship into a \textit{human-plus-technology} triad, introducing a second potential authority and raising the question of whether the peer or the model will be trusted as the primary interpreter of the situation. This shift can expand what peer specialists and service users are able to do together, yet it also changes how authority is signaled, how authenticity is judged, and how trust flows or fractures within the interaction.
 
Participants described both possibilities and vulnerabilities. A trusted peer specialist can serve as a bridge to an LLM. When peers disclose why and how they are using the model, obtain consent, and narrate their reasoning, service users often borrow trust from the human relationship and provisionally extend it to the tool. This triangulated trust can create a safe, judgment-free space in which service users explore their relationship with technology, asking questions, expressing fears, and learning at their own pace alongside someone who shares their experience. In this mode, LLMs can reduce generalized anxiety about AI by converting abstract concerns into concrete, collaborative practice.

At the same time, triangulation introduces new risks. When LLM outputs are treated as authoritative, service users may question whether they are receiving authentic peer support or AI-filtered guidance \cite{ai_advice}. Errors, decontextualized advice, or clinical language do not merely reduce confidence in the AI; they directly erode the peer’s relational authority that is the moral and practical center of peer support. Because what heals in PRO settings is the relationship itself, any disruption to that relationship is also a redistribution of power, credibility, and legitimacy---away from lived experience and toward algorithmic reasoning. If polished model language compromises the peer’s voice, or if decisions appear to originate from the system, the authority rooted in shared lived experience becomes less visible and less trusted. Critically, \textbf{\textit{trust transfer is reversible}}. Because service users often extend trust to the LLM through the peer, visible model failures such as inaccuracy, irrelevance, or inappropriate advice can back-propagate harm \cite{characterizing_ai_harms}. These failures undermine confidence in both the technology and the peer and strain the relational fabric that makes PRO support effective.

Participants offered a constructive alternative: use LLMs not as sources of truth, but as prompts for reflection and co-learning \cite{inspiration_human_llm}. When peer specialists and service users co-interpret model suggestions---probing assumptions, aligning outputs with lived experience, and deciding together what to keep, revise, or discard---the technology becomes a prompt for dialogue, not a substitute for it. In this framing, trust is not about AI reliability, but about ensuring that the peer-service user relationship remains the core site of meaning-making and authority. This co-interpretive stance requires clear provenance (what came from the model versus the peer), visible consent checkpoints, and defaults that require human confirmation. Here, LLMs do not deliver trust; they participate in its co-construction when mediated through judgment, explanation, and relational care.  

\subsection{From Human-in-the-Loop to Lived-Experiences-in-the-Loop in AI Design}

Our findings suggest that integrating LLMs into peer-run behavioral health services is not simply a matter of technical alignment or human oversight. It requires a deeper epistemic shift that centers lived experience as a guiding force across AI design, deployment, and evaluation. While existing HCI literature offers important insights into human-in-the-loop AI design \cite{shen2022public, yang2019human, nakao2022toward, schirner2013future}, such frameworks rarely interrogate which humans are in the loop, what forms of knowledge they hold, or how that knowledge is valued. In high-stakes, relational domains like peer support, the most relevant knowledge is often experiential, grounded in recovery, trauma, marginalization, and the everyday realities of navigating public systems. Human-in-the-loop protects workflow; \textbf{\textit{lived-experience-in-the-loop}} protects relational authority, identity, legitimacy, and mutuality. What is needed is not just human review but \textbf{\textit{experiential co-authorship}}, where peer specialists and service users actively shape the system's behavior, logic, and values \cite{auditing_human_expertise,expertise_ai}. 

Our participants envisioned precisely this model. To them, systems should begin from recovery-centered and person-centered goals, then build outward from local knowledge and resources. That means drawing on domain frameworks and community expertise, curating prompt libraries tied to the service context, aligning language and culture, and grounding guidance in holistic models such as the Eight Dimensions of Wellness \cite{swarbrick2006wellness}. Local, county-specific resources and peer or community supports should be easy to find and use. Equally important is avoiding canned defaults that push people to standardized routes like Social Security or police involvement when safer, community-led options exist \cite{llm_ethical_co_creator}. Resources should prioritize safety, avoid involuntary interventions such as forced hospitalizations, and utilize peer-support networks rooted in community knowledge and shared lived experience for more empathetic and actionable guidance. In this framing, community stakeholders become coauthors of prompt libraries and recovery-grounded templates, moving beyond user-centered design toward epistemic alignment through community-centered LLM fine-tuning and alignment. 

Designing for lived-experience-in-the-loop also reorients how we think about \textit{failure}. In conventional AI systems, failure is framed as an accuracy problem \cite{why_do_models_hallucinate}. But in peer support, failure is relational. It is a moment that can either damage bonds or be used to repair them. Clear reflective procedures such as acknowledging errors, explaining why a suggestion worked or failed, noticing conflicts with peer knowledge, correcting courses, continuing to build expertise, and re-centering the relationship can turn breakdowns into teachable moments \cite{shen2021everyday}. These are care practices, not just safety protocols. They imply that any LLM used in this setting must be accountable not only for what it outputs but for how its presence is felt and negotiated within human relationships \cite{ai_accountability}.

Centering lived experience in AI design also requires confronting tensions between usefulness and safety, not as a fixed tradeoff, but as a series of decisions shaped by the real-world risks, preferences, and constraints of those most affected \cite{ai_safety_utility}. In communities facing complex and intersecting social needs, privacy is a deeply contextual concern \cite{wu2025navigating, sleeper2019tough}. Rather than treating privacy as a technical constraint, systems must embed mechanisms for \textbf{\textit{contextual, granular, and ongoing consent}} that reflect people’s real circumstances and comfort levels. Here, lived experience is not just an input but the lens through which decisions about retention, access, and transparency should be made.

Finally, participants emphasized the need for guardrails that preserve peer autonomy. LLMs should scaffold---not replace---the core work of peer support \cite{ai_teacher_relationship}. Governance structures must treat peers not only as testers of model outputs, but as ongoing governors of relational and epistemic authority. Lived experience must constrain, direct, and limit what the model is allowed to do. This means delineating clear boundaries between what can be automated (e.g., documentation, information retrieval) and what must remain human (e.g., sensemaking, emotional resonance, trust-building). LLMs should be positioned not as experts or authorities, but as tools whose value depends on how well they support the judgment, identity, and community wisdom of those using them.

In centering lived-experience-in-the-loop, our study highlights that designing for high-stakes, community-led care requires more than technical fixes. It demands an \textit{epistemic shift}: from optimizing outputs to honoring relational processes, and from abstract alignment to grounded, co-authored systems. This shift protects the relational authority, credibility, and meaning-making that make peer support effective and offers a model for how AI can remain accountable to the people and practices it aims to serve.

\section{Limitations and Future Work}
This study has several limitations. First, we did not formally evaluate the effectiveness of comicboarding as a method. As a formative study aimed at understanding how PROs conceptualize and govern LLMs rather than evaluating a deployed system, we used comicboards to support early design exploration and help participants reason about potential LLM-mediated interactions rather than assessing the method’s efficacy. Future work should incorporate mixed-methods measures and controlled evaluations of LLM comicboarding to clarify when and how it elicits useful feedback compared with interviews or other co-design methods. Such work would strengthen assessments of understanding, engagement, and benefit-risk reflection, especially for participants with diverse literacy levels. Second, because comicboarding supports early exploration without requiring system implementation, our workshops lacked hands-on interaction with an actual LLM, which may have limited participants' ability to assess feasibility, usability, or real-world harms. Future research should integrate live demonstrations or in situ use of existing tools to improve ecological validity. Third, our workshops occurred after CSPNJ had already committed to exploring LLM adoption. While comicboarding supported meaningful feedback on design directions, using the method earlier could surface more foundational concerns, including whether LLMs are appropriate at all. Fourth, because our comicboards intentionally abstracted implementation details to avoid prescriptive interfaces, participants may have been limited in envisioning concrete advantages or critiquing interface-level design. Future work might pair comicboarding with interactive or high-fidelity prototypes to surface stakeholders’ expectations and concerns more fully at the interaction level. Fifth, while some stakeholder-informed strategies such as RAG, prompt libraries, and community fine-tuning are technically feasible today, others remain aspirational and require further research, particularly in hallucination detection and contextual sensitivity. We view these strategies not as plug-and-play solutions but as design requirements and guiding values that should shape system development and deployment over time. Finally, our findings are grounded in a single peer-run behavioral health organization in one U.S. state. While the insights may generalize to other resource-constrained, community-led service settings, future work should examine how these methods and themes apply across additional domains, populations, and AI use cases.
\section{Conclusion}

In this study, conducted in partnership with CSPNJ, we contribute an empirical understanding of how peer specialists and service users perceive the opportunities, risks, and mitigation strategies associated with integrating LLMs into PROs. Using comicboarding to expand stakeholder participation, we elicited feedback from individuals who often face structural barriers to engagement and surfaced the values and lived experiences shaping peer support practice. Participants described LLMs as potentially valuable collaborators---tools that may scaffold relational work without displacing its human foundation---while also identifying risks of contextual misalignment, relational erosion, and the displacement of experiential peer authority. They further articulated actionable strategies for aligning systems with lived experience, community resources, and recovery-oriented care. Our findings extend HCI literature by advancing a shift from human-in-the-loop toward lived-experience-in-the-loop design. We argue that responsible AI in relational, high-stakes settings must center experiential knowledge across system design, behavior, governance, and accountability to protect the relational authority through which support, trust, and meaning are co-constructed in community-led care.

\begin{acks}
This research was supported by the National Science Foundation (NSF) Civic Innovation Challenge (CIVIC) program under Award Nos. 2527408 and 2431230. We thank all of our collaborators at Collaborative Support Programs of New Jersey (CSPNJ) for their invaluable feedback, insights, ongoing collaboration, and sustained partnership throughout this project. We also thank our anonymous reviewers for their thoughtful and constructive feedback in improving this work.
\end{acks}

\bibliographystyle{ACM-Reference-Format}
\bibliography{citations}

%\newpage
%\clearpage
\appendix
\label{sec:appendix}

\section{LLM-Lifecycle Comicboards}

We include all the comicboards we developed and used for our study. The comicboards were designed as sequential visual narratives that translate technical LLM processes into concrete, situated scenarios drawn from peer specialists’ everyday workflows. Across the appendix, the comicboards are organized to reflect different stages of LLM use, including an introductory overview of what LLMs are, followed by stage-based depictions of what the LLM does during sessions, what inputs it requires, how it improves over time, and our proposed solutions for how risks are mitigated across stages. Each comicboard combines simplified illustrations, short captions, and guiding questions to prompt reflection and discussion on information flow, opportunities, risks, mitigation strategies, and overall feedback. This method enabled our participants to critically engage with LLM behavior, surface contextual knowledge that might otherwise remain implicit, and collaboratively articulate expectations for responsible and context-aware AI use in peer-run services.

\setlength{\floatsep}{3\baselineskip}

\begin{figure*}[!htb]
    \centering
    \includegraphics[width=\textwidth]{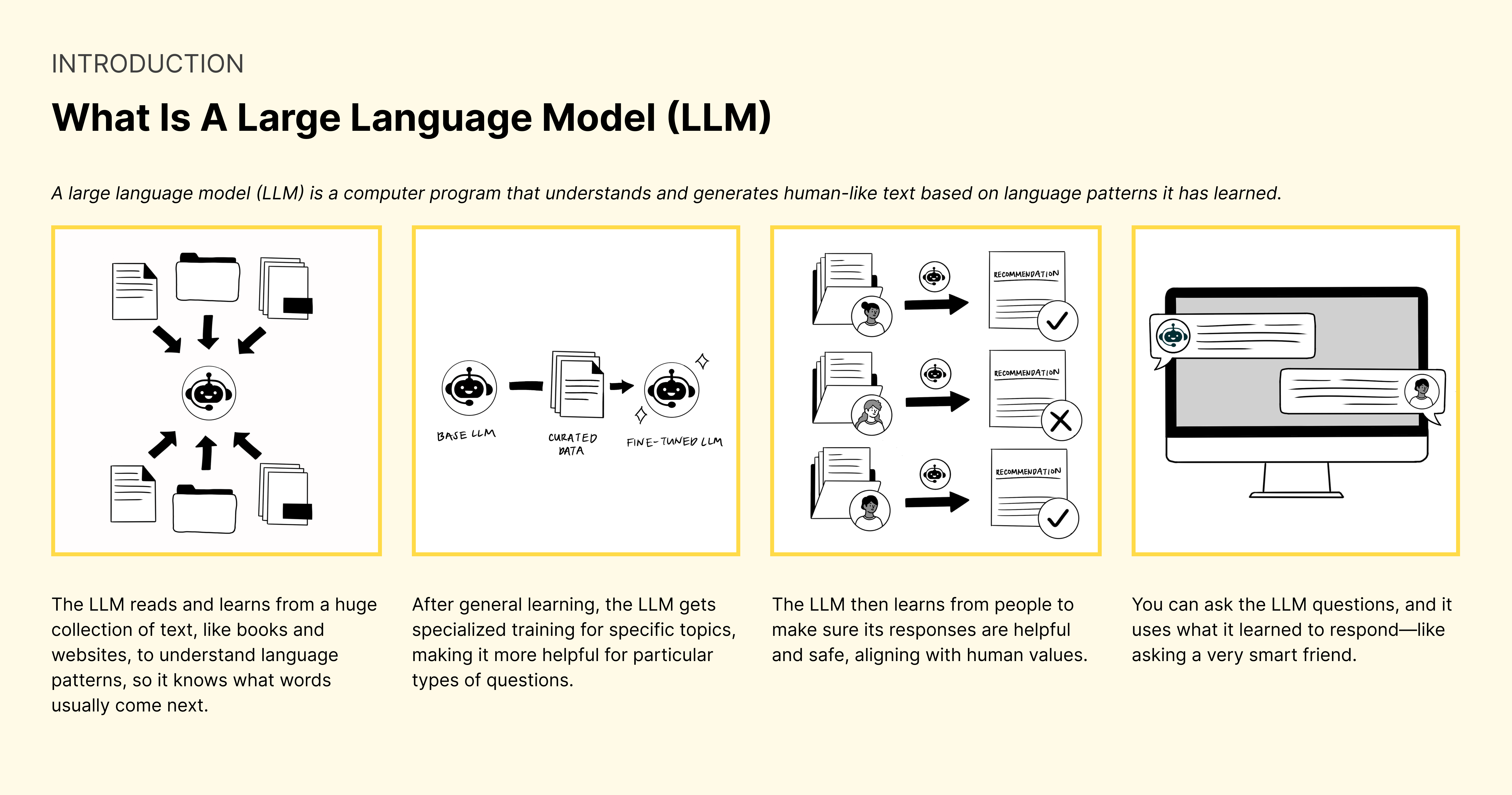}
    \caption{Introduction comicboards (“What is an LLM?”) illustrating the definition of LLMs and four key aspects of the LLM development process: pre-training, supervised fine-tuning, reinforcement learning, and prompting.}
    \Description{A four-panel comicboard introducing large language models. Panel one shows many documents and websites feeding into an LLM to illustrate large-scale text pre-training. Panel two shows the LLM receiving specialized training for specific tasks. Panel three depicts people providing feedback to align the LLM’s responses with human values. Panel four shows a person interacting with the LLM on a computer, representing asking questions and receiving responses.}
    \label{fig:comicboard0}
\end{figure*}

\begin{figure*}[!htb]
    \centering
    \includegraphics[width=\textwidth]{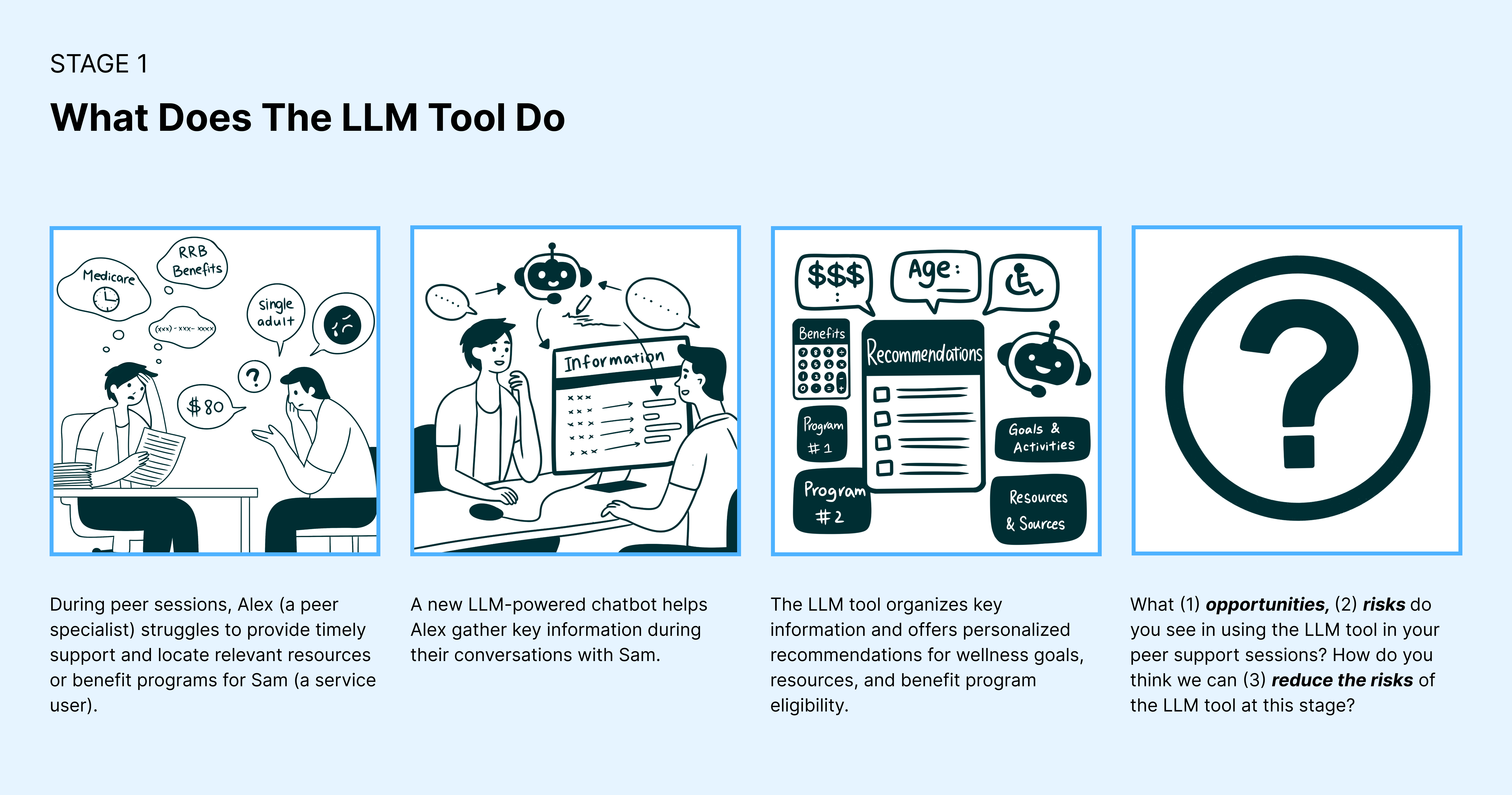}
    \caption{Stage 1 comicboards (“What does the LLM tool do?”) depicting how an LLM assists peer specialists in gathering key information during sessions, organizing it, and providing personalized recommendations for wellness goals, resources, and benefit eligibility, to support timely and context-aware peer support in peer-run behavioral health settings.}
    \Description{A four-panel comicboard depicting how an LLM is used during peer support sessions. Panel one shows a peer specialist and service user discussing needs such as benefits and resources. Panel two shows the peer specialist using an LLM tool to gather and organize information during the conversation. Panel three shows the LLM presenting personalized recommendations for wellness goals, resources, and benefit eligibility. Panel four contains a question prompt inviting reflection on opportunities, risks, and risk reduction or mitigation strategies at this stage.}
    \label{fig:comicboard1}
\end{figure*}

\begin{figure*}[!htb]
    \centering
    \includegraphics[width=\textwidth]{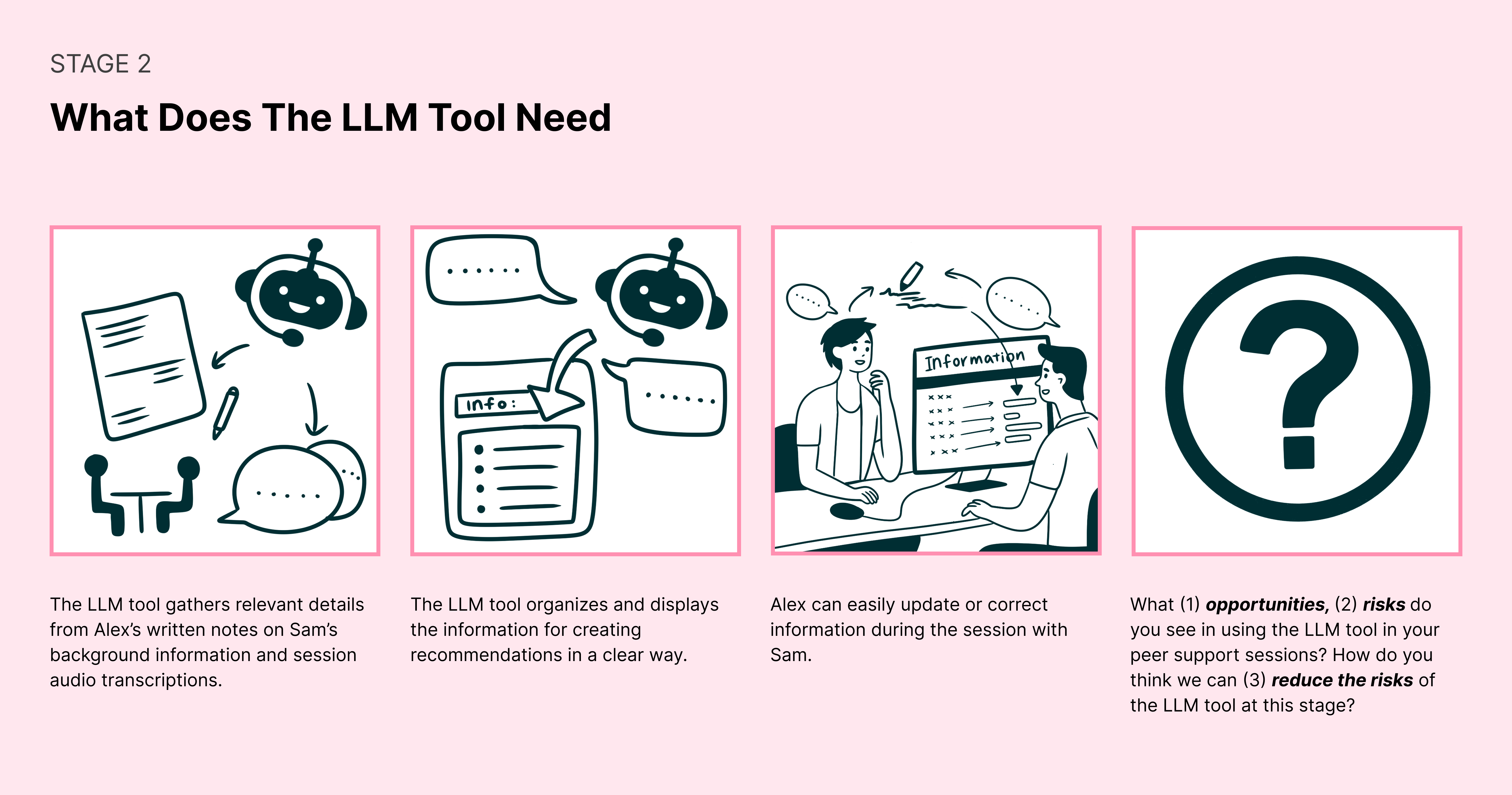}
    \caption{Stage 2 comicboards (“What does the LLM tool need?”) illustrating how an LLM gathers and organizes information from peer specialists’ notes and session transcripts, displays recommendations clearly, and enables real-time updates or corrections in peer-run behavioral health settings.}
    \Description{A four-panel comicboard illustrating what information the LLM requires to function. Panel one shows the LLM receiving notes and audio transcripts from peer support sessions. Panel two depicts the LLM organizing and displaying information clearly for recommendation-making. Panel three shows the peer specialist updating or correcting information in real time during a session. Panel four contains a question prompt inviting reflection on opportunities, risks, and risk reduction or mitigation strategies at this stage.}
    \label{fig:comicboard2}
\end{figure*}

\begin{figure*}[!htb]
    \centering
    \includegraphics[width=\textwidth]{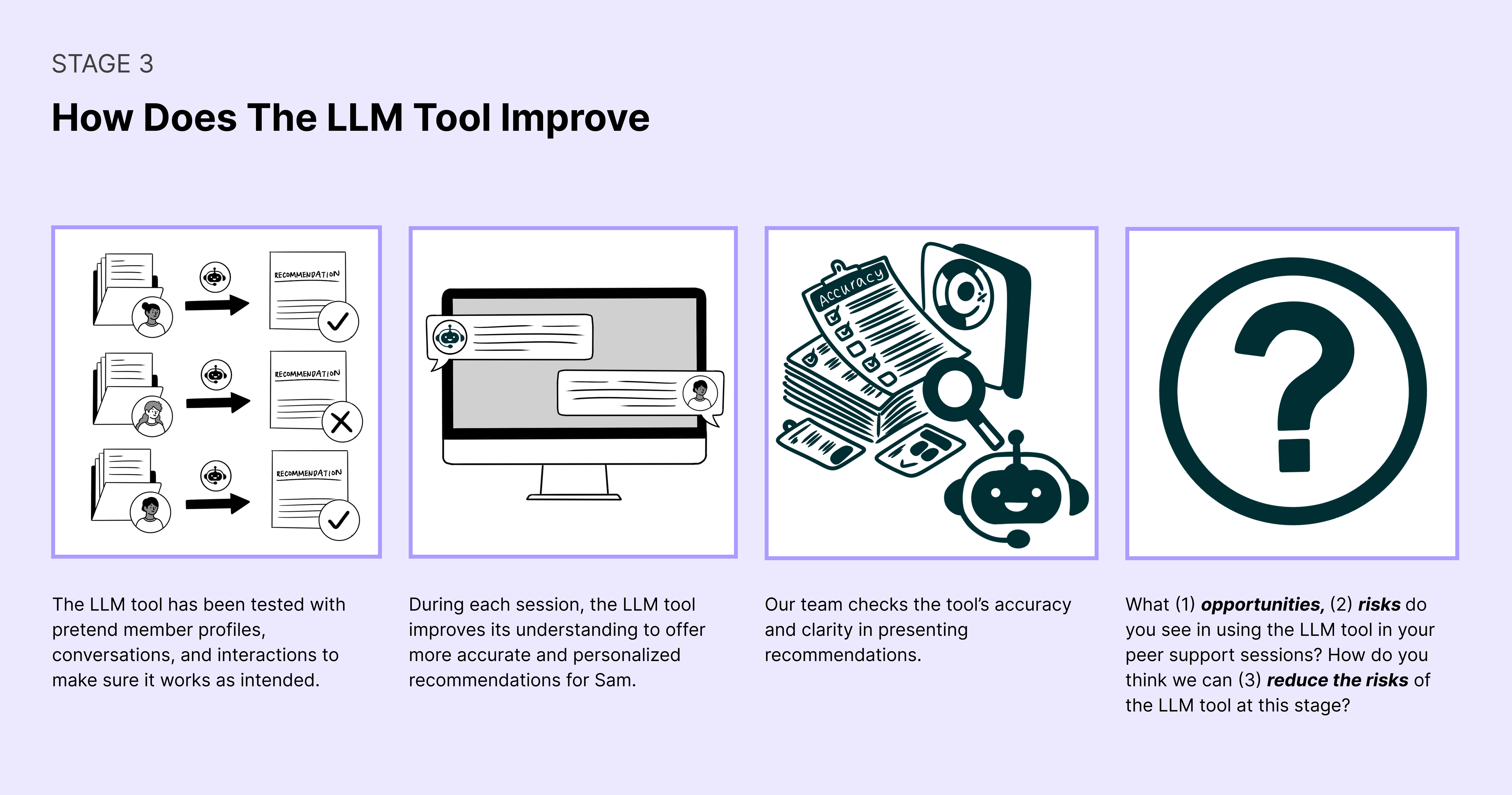}
    \caption{Stage 3 comicboards (“How does the LLM tool improve?”) depicting iterative system improvement through testing with representative profiles, in-session learning for more accurate and personalized recommendations, and team-based audits.}
    \Description{A four-panel comicboard showing how the LLM improves over time. Panel one depicts testing with representative user profiles and simulated interactions. Panel two shows the LLM learning during sessions to produce more accurate and personalized recommendations. Panel three shows a team reviewing outputs through audits and evaluation materials. Panel four contains a question prompt inviting reflection on opportunities, risks, and risk reduction or mitigation strategies at this stage.}
    \label{fig:comicboard3}
\end{figure*}

\begin{figure*}[!htb]
    \centering
    \includegraphics[width=\textwidth]{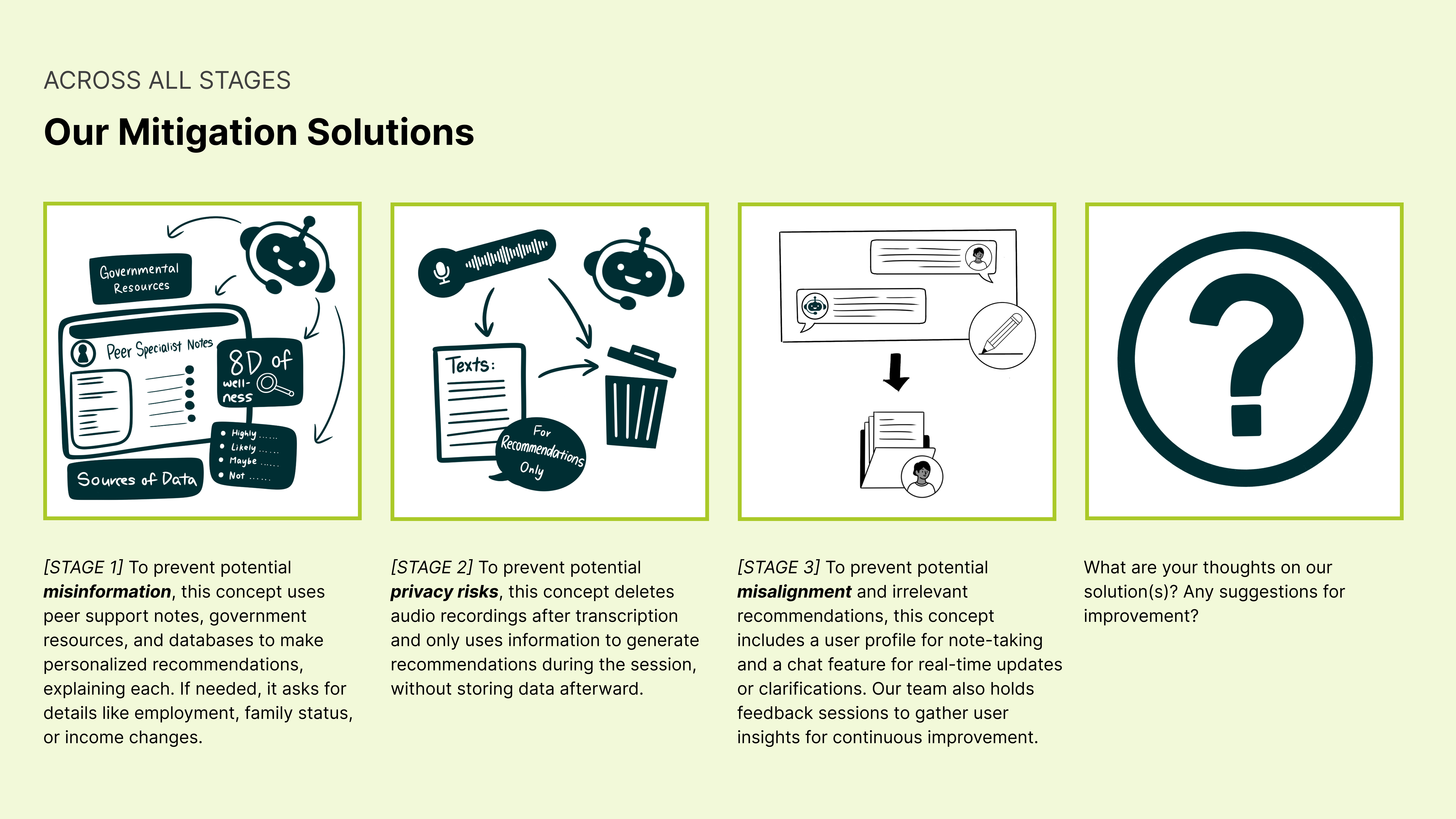}
    \caption{Across-all-stages comicboards (“Our mitigation solutions”) illustrating the research team's mitigation solutions, showing how user alignment, data minimization (e.g., deleting audio after transcription), transparent user profiles, real-time updates, and continuous feedback sessions collectively address misinformation, privacy risks, and misalignment to ensure LLMs remain safe, context-aware, and responsive to community needs in peer-run behavioral health settings.}
    \Description{A four-panel comicboard presenting mitigation strategies applied across all stages of LLM use. Panel one shows curated data sources and alignment with user needs to prevent misinformation. Panel two depicts data minimization practices, including deleting audio after transcription. Panel three shows transparent user profiles, real-time updates, and feedback sessions for continuous improvement. Panel four contains a question prompt inviting feedback and suggestions on the proposed mitigation solutions.}
    \label{fig:comicboard4}
\end{figure*}

\end{document}